\documentclass[prx,aps,longbibliography,showpacs,groupedaddress,superscriptaddress,twocolumn,toc=flat,nofootinbib]{revtex4-1}

\usepackage{graphicx}
\usepackage{latexsym}
\usepackage{amssymb}
\usepackage{amsmath}
\usepackage{amsfonts}
\usepackage{makecell}
\usepackage{enumerate}
\usepackage{upgreek}
\usepackage{subfigure} 
\usepackage{bm}
\usepackage{nicefrac}
\usepackage{bbold}
\usepackage{verbatim}
\usepackage{multirow}
\usepackage{color}
\usepackage{comment}
\usepackage{xcolor}
\usepackage{soul}
\usepackage{physics}
\usepackage{tikz}
\usepackage{braket}
\usepackage{ragged2e} 
\usepackage{amsmath}
\usepackage{xcolor}
\usepackage{dsfont}
\usepackage{mathtools}
\usepackage{marvosym}
\usepackage{glossaries}

\usepackage[hidelinks,backref=false,unicode=true,breaklinks=false, bookmarks=false,colorlinks=true]{hyperref}

\usepackage{cleveref}

\AtBeginDocument{%
    \newwrite\bibnotes
    \def\bibnotesext{Notes.bib}
    \immediate\openout\bibnotes=\jobname\bibnotesext
    \immediate\write\bibnotes{@CONTROL{REVTEX41Control}}
    \immediate\write\bibnotes{@CONTROL{%
    apsrev41Control,author="08",editor="1",pages="1",title="0",year="1"}}
     \if@filesw
     \immediate\write\@auxout{\string\citation{apsrev41Control}}%
    \fi
}%

\newacronym{HO}{HO}{hidden order}
\newacronym{HOP}{HOP}{hidden order phase}
\newacronym{DOF}{DOF}{degree of freedom}
\newacronym{AFM}{AFM}{antiferromagnetic}
\newacronym{PM}{PM}{paramagnetic}
\newacronym{SSB}{SSB}{spontaneous symmetry breaking}
\newacronym{LRO}{LRO}{long-range order}
\newacronym{SPT}{SPT}{symmetry protected topological}
\newacronym{NN}{NN}{nearest-neighbor}
\newacronym{NNN}{NNN}{next-nearest-neighbor}
\newacronym{QCP}{QCP}{quantum critical point}
\newacronym{hQCP}{hQCP}{hidden quantum critical point}
\newacronym{TFIM}{TFIM}{transverse field Ising model}
\newacronym{RIXS}{RIXS}{resonant inelastic X-Ray scattering}

\newacronym{DMRG}{DMRG}{density matrix renormalization group}
\newacronym{MPS}{MPS}{matrix-product state}
\newacronym{MPO}{MPO}{matrix-product operator}
\newacronym{LSE-TDVP}{LSE-TDVP}{local subspace expansion time-dependent variational principle}
\newacronym{TDVP}{TDVP}{time-dependent variational principle}
\newacronym{CTKS}{CTKS}{complex-time Krylov space}
\newacronym{DSF}{DSF}{dynamical spin-structure factor}
\newacronym{BZ}{BZ}{Brillouin zone}

\newcommand{\ITAMP}{ITAMP, Harvard-Smithsonian Center for Astrophysics, Cambridge, MA 02138, USA}
\newcommand{\Harvard}{Department of Physics, Harvard University, Cambridge, MA 02138, USA}

\usepackage[german,english]{babel}
\hypersetup{
    colorlinks=true,
    linkcolor=blue!60!black,
    citecolor=blue!60!black,
    urlcolor=blue!60!black
}

\begin{document}
\title{Spectral signatures of hidden order and hidden quantum critical points}

\author{Reja H. Wilke}
\affiliation{Institute for Theoretical Physics, ETH Z\"urich, CH-8093 Zurich, Switzerland}

\author{Henning Schl\"omer}
\affiliation{\ITAMP}
\affiliation{\Harvard}

\author{Sebastian Paeckel}
\affiliation{Department of Physics and Arnold Sommerfeld Center for Theoretical Physics (ASC), Ludwig-Maximilians-Universit\"at M\"unchen, Theresienstr. 37, M\"unchen D-80333, Germany}
\affiliation{Munich Center for Quantum Science and Technology (MCQST), Schellingstr. 4, M\"unchen D-80799, Germany}

\author{Annabelle Bohrdt}
\affiliation{Department of Physics and Arnold Sommerfeld Center for Theoretical Physics (ASC), Ludwig-Maximilians-Universit\"at M\"unchen, Theresienstr. 37, M\"unchen D-80333, Germany}
\affiliation{Munich Center for Quantum Science and Technology (MCQST), Schellingstr. 4, M\"unchen D-80799, Germany}

\author{Fabian Grusdt}
\affiliation{Department of Physics and Arnold Sommerfeld Center for Theoretical Physics (ASC), Ludwig-Maximilians-Universit\"at M\"unchen, Theresienstr. 37, M\"unchen D-80333, Germany}
\affiliation{Munich Center for Quantum Science and Technology (MCQST), Schellingstr. 4, M\"unchen D-80799, Germany}

\date{\today}

\begin{abstract}
Hidden-order phases elude the conventional Ginzburg-Landau framework: although they spontaneously break symmetry, their order remains invisible to standard local two-point correlators. Instead, they are typically characterized by nonlocal (string) order parameters and are believed to underlie a range of exotic quantum states, such as symmetry-protected topological (SPT) phases. While hidden correlations have previously been detected through nonlocal observables in quantum gas microscope experiments, in realistic solid-state systems multipoint observables are typically inaccessible and the signatures of hidden order remain largely unexplored. In this work, we show that hidden order in doped spin chains can be directly probed through dynamical two-point response functions measurable by resonant inelastic X-ray scattering (RIXS) and neutron scattering. Focusing on a minimal model, we numerically identify clear spectral fingerprints of both the hidden-order phase, its associated hidden quantum critical point as well as a specific group of gapless SPT phases. Complementing our numerical results with analytical insights, we uncover the microscopic origin of these spectral signatures and demonstrate how they emerge from the interplay between spin and charge degrees of freedom. Together, our results establish spectroscopy as a practical probe of hidden order and gapless SPT phases and provide a unified understanding of the low-energy physics governing these unconventional quantum phases.
\end{abstract}
\maketitle

Hidden order~\cite{kruis_geometry_2004, wilke_topology_2026} constitutes a form of quantum order beyond the conventional Landau paradigm. While hidden order phases lack a local order parameter and conventional \gls{LRO}, they are defined by a nonlocal realization of a spontaneously broken symmetry and the corresponding ground-state degeneracy. This structure reveals itself through nonlocal (string) order parameters, while conventional two-point functions remain short-ranged even as the system resides in a nontrivial ordered phase.

\begin{figure}
\includegraphics[width=.5\textwidth]{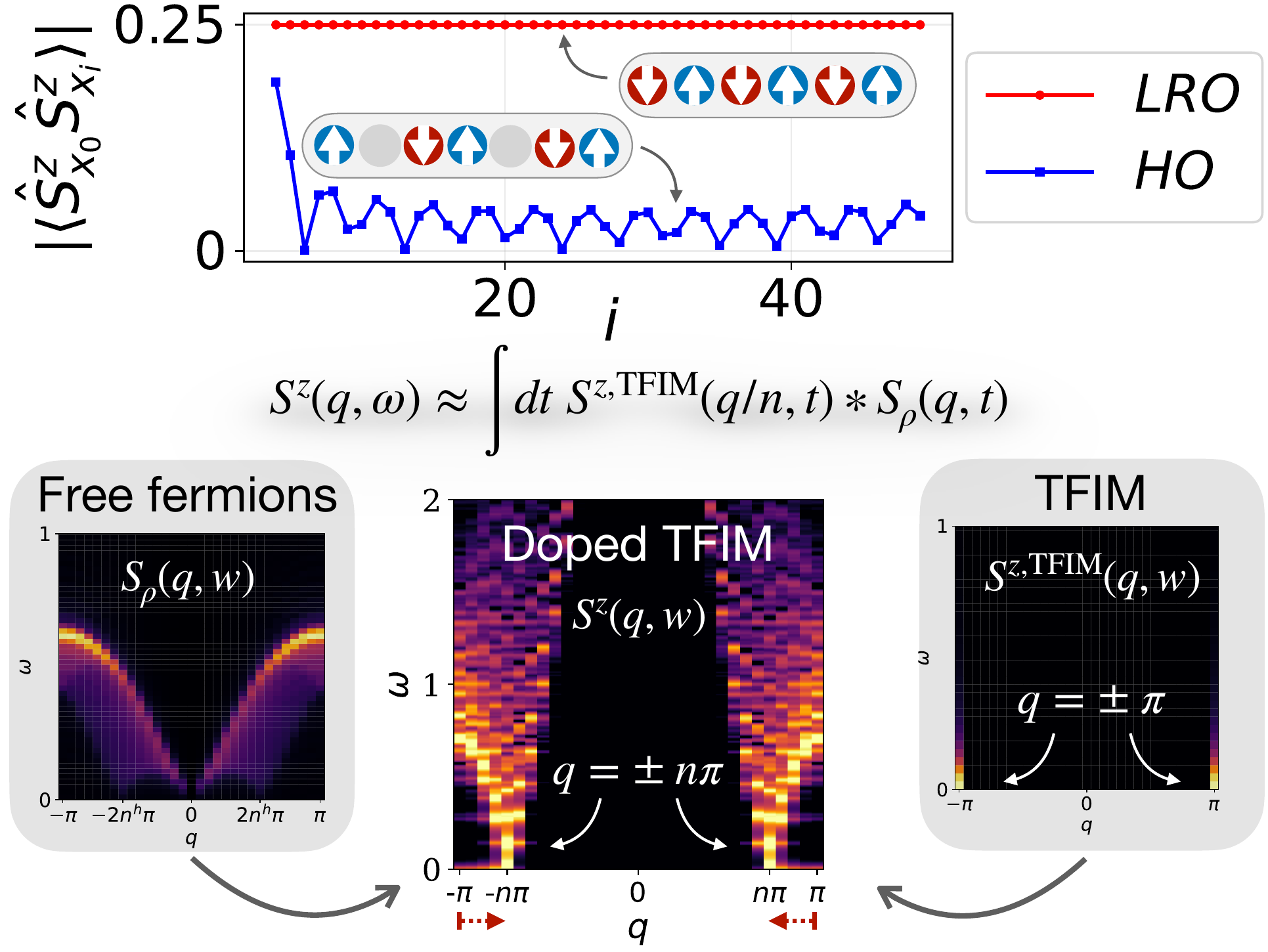}
\caption{\justifying Spectral signatures of hidden order. Hidden order (\gls{HO}) phases are characterized by a nonlocal realization of a spontaneously broken symmetry. Due to their nonlocal character, they cannot be detected via conventional equal-time two-point correlation functions, as is the case for antiferromagnetic long-range order (\gls{LRO}). Upper panel: Comparison of the equal-time two-point correlator $\vert\hat S^z_{x_0} \hat S^z_{x_i}\vert$ in a~\gls{HO} phase (blue curve) and an antiferromagnetic~\gls{LRO} phase (red curve). Insets show exemplary spin configurations in the respective phases. Lower panel: Sketch of our approach. We probe~\gls{HO} through its spectral fingerprints obtained via \textit{dynamical} two-point correlators, see Fig.~\ref{fig:dsf} and Fig.~\ref{fig:dsf_x}. We show that the dynamical structure factor of the doped~\gls{TFIM} $S^{z}(q,\omega)$ (center) is given by the convolution of the charge structure factor of a free fermion model $S_\rho (q,t)$, eq.~\eqref{eq:Hc}, and the spin structure factor of the conventional~\gls{TFIM} $S^{z, \rm TFIM}(q/n,t)$, eq.~\eqref{eq:Hs}, see~\cref{sec:analytical_sf}. We analyze the effect of doping on the spectral features, such as a shift in the gapless points from $q=\pm\pi$ (TFIM) to $\rightarrow q=\pm n\pi$ (doped TFIM).}\label{fig:intro}
\end{figure}

A paradigmatic example is provided by the spin-1 Heisenberg chain~\cite{den_nijs_preroughening_1989, kennedy_hidden_1992}, where antiferromagnetic correlations are obscured by intervening spin-$0$ states. In recent years, hidden order has been identified as the driving mechanism behind different exotic quantum phases~\cite{read_order_1989, girvin_off-diagonal_1987, thorngren_intrinsically_2021,pauw_detecting_2024}. In particular, concepts describing hidden order -- mostly relying on string order parameters -- were reinterpreted within the framework of~\gls{SPT} phases~\cite{pollmann_detection_2012, else_hidden_2013, thorngren_intrinsically_2021, wilke_topology_2026}. 
Recent advances in quantum simulation have demonstrated that hidden-order correlations can be accessed experimentally through nonlocal observables and string correlators~\cite{hilker_revealing_2017}. In realistic materials, the experimental characterization of hidden order remains far less developed. Equal-time two-point correlation functions alone provide only limited information, while the required multipoint observables are generally difficult to access in solid-state settings. This motivates the search for experimentally accessible signatures of hidden order in conventional response functions. In this work, we therefore investigate how hidden order can be probed through spectroscopic observables and \emph{dynamical} two-point correlation functions, which are directly measurable in real materials and widely available in condensed-matter experiments. For an overview of our approach see Fig.~\ref{fig:intro}.

In the present article, we study one of the conceptually simplest realizations of hidden order: a doped~\gls{TFIM}, that hosts a gapless topological phase~\cite{thorngren_intrinsically_2021, keselman_gapless_2015, scaffidi_gapless_2017, parker_topological_2018}, characterized by a non-trivial string order and spin-$1/2$ edge modes.
We investigate signatures of~\gls{HO} in the dynamical spin structure factor, which can be probed as a function of momentum and energy using~\gls{RIXS} or neutron scattering experiments~\cite{ament_resonant_2011, lynn_magnetic_2012}. As a momentum- and energy-resolved two-point response function, the dynamical structure factor encodes a wealth of information about the hidden magnetic order of the system -- insights that cannot be accessed through local static probes alone. Using matrix product states, we uncover remarkably rich spectral signatures of hidden magnetic order, demonstrating that hidden-order phases can be identified through their characteristic dynamical response.

We rationalize our numerical findings through complementary analytical considerations, confirming that doping -- the essential ingredient for hidden order in this setting -- primarily induces a renormalization of the spectral features already present in the undoped system. Our analytical approach is based on a squeezed space construction~\cite{ogata_bethe-ansatz_1990, kruis_geometry_2004}, which approximately decouples spin and charge degrees of freedom in the model.
This decomposition provides a transparent physical interpretation of the spectral response in terms of the underlying spin and charge excitations.\\

Furthermore, the investigated~\gls{HO} phase can be understood as an \textit{intrinsically gapless} topological phase~\cite{thorngren_intrinsically_2021}. In contrast to traditional symmetry-protected topological (SPT) phases, which rely on a robust bulk energy gap, the topological characteristics of this phase -- its non-trivial string order and localized spin-1/2 edge modes -- strictly require the bulk to remain gapless. This absence of a bulk energy gap naturally raises the question of the spectral signatures of such phases. We show that indeed, signatures of this symmetry-protection can be observed in the excitation spectrum.

Our findings of long-lived, spin-wave like quasi-particle excitations with a renormalized dispersion are reminiscent of paramagnons discovered in cuprate superconductors~\cite{xu_testing_2009,le_tacon_intense_2011}. Paramagnons occur in the absence of \gls{LRO} and have been demonstrated to exist in the pseudogap phase~\cite{xu_testing_2009, jia_persistent_2014, dean_persistence_2013} in \gls{RIXS}. The coexistence of hidden order and paramagnons in the pseudogap phase has been theoretically explored in~\cite{schlomer_geometric_2025}.

The paper is organized as follows: In~\cref{sec:model}, we introduce the $t_1-t_2-$~\gls{TFIM} and discuss the detection of hidden order via nonlocal order parameters. Section~\ref{sec:results} presents our main findings, beginning with the phase diagram of the model in~\cref{sec:phase_diagram}. Subsequently,~\cref{sec:numerical_sf} and~\cref{sec:analytical_sf} detail the~\gls{DSF} derived in independent numerical and analytical approaches. Finally, we combine both techniques to draw systematic conclusions on the effect of doping on the excitation spectrum in the longitudinal~\gls{DSF},~\cref{sec:HO_discussion}, before analyzing the spectral signatures of hidden order in the transverse~\gls{DSF},~\cref{sec:spectral_signatures_discussion}.

\section{Model\label{sec:model}}
For our investigation of hidden order, we choose a minimal model that exhibits a strong interplay between spin and charge degrees of freedom: The~\gls{AFM} $t_1-t_2-$ \gls{TFIM}, a doped version of the~\gls{AFM}~\gls{TFIM}, which features a long-range ordered, a paramagnetic and a hidden order phase.
The Hamiltonian of the system under consideration is given by
\begin{equation}
\begin{aligned}
		\hat{H} = &-t_1 \sum_{\langle i,j\rangle, \sigma} \hat{\mathcal{P}}_{GW}(\hat c^\dagger_{i\sigma}\hat c^{\phantom\dagger}_{j\sigma} + \mathrm{h.c.}) \hat{\mathcal{P}}_{GW}\\
        &- t_2 \sum_{\langle\langle i,j\rangle\rangle, \sigma} \hat{\mathcal{P}}_{GW}(\hat c^\dagger_{i\sigma}\hat c^{\phantom\dagger}_{j\sigma} + \mathrm{h.c.}) \hat{\mathcal{P}}_{GW}\\
		&+ J \sum_{\langle i,j\rangle} \hat S^z_i  \hat S^z_j + h_x \sum_j \hat S^x_j \;.
		\label{eq:doped_hopping_TFIM}
\end{aligned}
\end{equation}
Here, the annihilation (creation) operator $\hat c^{(\dagger)}_{j\sigma}$ annihilates (creates) a fermion with spin $\sigma = \uparrow,\downarrow$ on site $j$.
The local density operator is given by $\hat n_j = \hat c^{\dagger}_{j\sigma}\hat c_{j\sigma}$ and the hole operator by $\hat n^h_j = \hat{\mathds{1}}-\hat n_j$ with respective eigenvalues $n^{(h)}_j$.
The model exhibits \gls{NN} and \gls{NNN} hopping with amplitude $t_1$ and $t_2$, respectively. The coupling strength of the Ising interaction is denoted by $J$ and the transverse field strength by $h_x$. The Gutzwiller projectors $\hat{\mathcal{P}}_{GW}$ remove double occupations, i.e. we work in a $t-J$ basis.
Furthermore, the model has a global $\mathbb{Z}_2$ spin symmetry, $\hat S^x_{i} \rightarrow \hat S^x_{i}, \hat S^y_{i} \rightarrow -\hat S^y_{i}, \hat S^z_{i} \rightarrow -\hat S^z_{i}$ for all $i$. Unless stated differently, we consider systems of size $L=30$ doped with $\hat N^h = \sum_{j=1}^L \hat n^h_j = 8$ holes.

The case of $t_2 = 0$ will be referred to as the $t_1$-\gls{TFIM}, which reduces to a $t-J_z$ model when $h_x=0$ and has previously been discussed by Thorngren et al.~\cite{thorngren_intrinsically_2021}. Their investigations first identified the hidden order phase via its non-trivial string order and edge modes, categorizing it as an~\textit{intrinsically gapless} topological phase. Crucially, this hidden order is dynamically stabilized by a gapless Luttinger liquid of doped holes, whose continuous charge fluctuations disrupt the local spin background and suppress long-range magnetic order. If the charge carriers were to localize, the system would freeze into an antiferromagnetic state, breaking the protecting symmetry and annihilating the topological edge modes. Therefore, the gaplessness of the charge fluid is essential: it prevents spontaneous symmetry breaking, preserves the global symmetry, and protects the non-local topological string order within the bulk.
Furthermore, Thorngren et al.~\cite{thorngren_intrinsically_2021} showed that the existence of this specific string correlator -- taking the form $\langle \hat S^z_i (\prod_{i<k<j} \hat P_k) \hat S^z_j \rangle$, where $\hat P_k$ is the local fermion parity operator applied along the string -- strictly prohibits the opening of a bulk gap, relying on the gapless nature of the Luttinger liquid to stabilize the edge states.

\begin{figure}
\includegraphics[width=.38\textwidth]{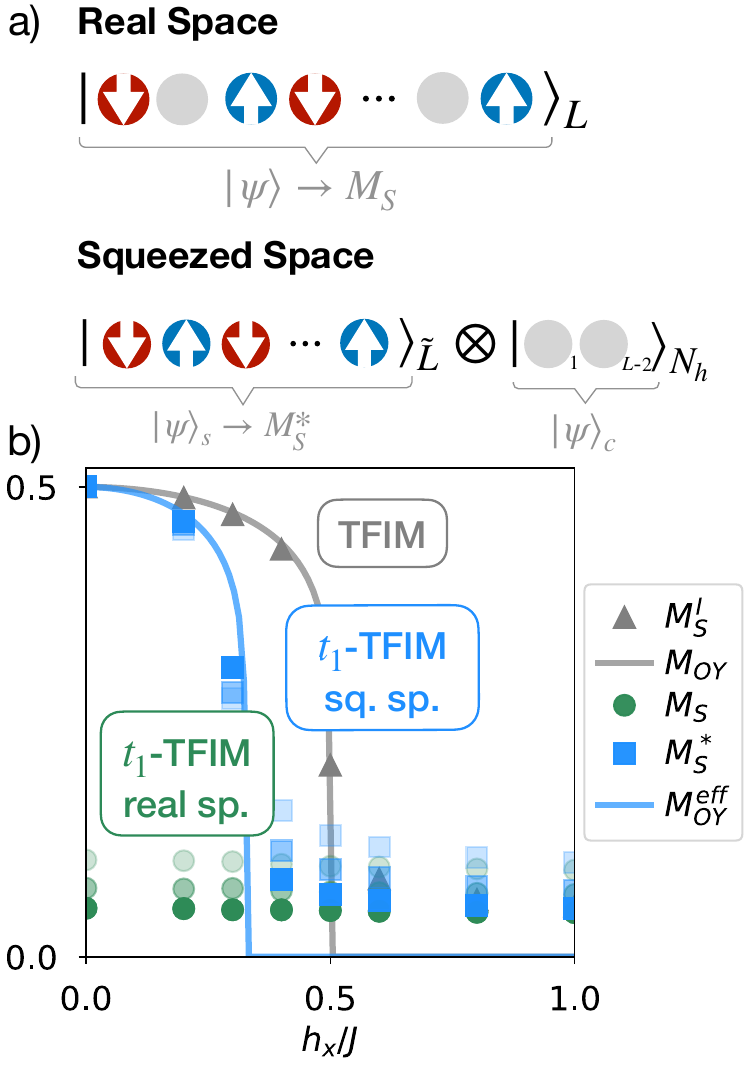}
    \caption{\justifying Detecting hidden order in squeezed space.
    $a)$ Illustration of a spin configuration in real and squeezed space, used to compute the corresponding staggered magnetizations, $M_S$ and $M_S^*$ (see~\cref{sec:model}).
    $b)$ Staggered magnetizations in real and squeezed space for the $t_1$-\gls{TFIM} as a function of $h_x$ at $t_1/J = 1$. The squeezed-space staggered magnetization $M_S^*$ (blue squares) remains finite for $h_x \leq 0.3 J$, indicating the presence of antiferromagnetic long-range order in squeezed space, referred to as hidden order. By contrast, the real-space staggered magnetization $M_S$ (green circles) does not detect this order.
    For reference, we also show the real-space staggered magnetization of the conventional~\gls{TFIM}, $M^I_S$ (gray triangles), which exhibits long-range order for $h_x \leq 0.5 J$.
    The Onsager-Yang prediction, $M_{OY}(J)=\frac{1}{2}\left[1-\left(\frac{2h_x}{J}\right)^2\right]^{1/8}$ (gray line), and the corresponding effective prediction $M^{\rm{eff}}_{OY}(J^{\rm snap}_{\rm eff})$ (blue line), agree well with the numerical results. The effective coupling $J^{\rm snap}_{\rm eff}\approx0.65$ is determined numerically via~\cref{eq:Jeff}.
    Increasing opacity denotes increasing system size, $L=30,60,120$, at fixed filling $n=N/L=22/30$. All observables are computed from $z$-basis snapshots of the many-body wavefunction obtained using~\gls{DMRG}.}
    \label{fig:sqsp} 
\end{figure}

\emph{Squeezed space.--}
While most characterizations of hidden order feature string correlators, in the strong coupling limit, $U/t \rightarrow \infty$, of the Fermi-Hubbard model, squeezed space provides an alternative route:
As shown by Ogata and Shiba~\cite{ogata_bethe-ansatz_1990}, here the ground state factorizes exactly into independent spin and charge components:
The charge degrees of freedom are expressed as a Slater determinant of spinless fermions, while its spin degrees of freedom are equivalent to the one-dimensional spin-$\frac{1}{2}$ Heisenberg model, which is solvable by Bethe Ansatz~\cite{bethe_zur_1931}.
Note that this is the case for arbitrary doping.
Central to this construction is that the spins are not labeled on the original lattice but instead on an effective lattice with all empty sites removed.
This observation gave rise to the squeezed-space formalism for describing doped spin chains in one dimension~\cite{kruis_geometry_2004, hilker_revealing_2017}, a framework that carries over to more general spin systems~\cite{grusdt_z2_2020, schlomer_robust_2023} and beyond the exactly soluble limit.
More specifically, in squeezed space, the ground state wave function $\vert \psi\rangle =\vert \sigma_1, \dots, \sigma_L\rangle$ with $\sigma_j = \uparrow, \downarrow, h$ is relabeled as $\vert \psi\rangle = \vert \tilde\sigma_1, \dots, \tilde \sigma_{\tilde L}\rangle \otimes \hat h^\dagger_{x_1}\dots \hat h^\dagger_{x_{N^h}} \vert 0\rangle$, with $\tilde \sigma_j = \uparrow, \downarrow$ and $\tilde L = L -N^h$.
Here, $N^h$ denotes the number of holes and $\hat h^\dagger_j$ corresponds to the creation operator of a hard-core chargon with the same statistics as $\hat c_{j,\sigma}$.
After this relabeling, the wave function can be decoupled into individual spin and charge wave functions~$\vert \psi\rangle =\vert \psi\rangle_s \otimes\vert \psi\rangle_c$, see Fig.~\ref{fig:sqsp} a).
The location of a spin in squeezed space is related to a spin in real space via $\tilde j = j - \sum_{l<j} n^h_l$.
Furthermore, the spin interaction in squeezed space experiences a density-dependent renormalization~\cite{hilker_revealing_2017}
\begin{equation}
    J_{\rm eff} = \frac{J}{L-1} \sum_{i=1}^{L-1}\frac{\langle (1-\hat n^h_i)(1-\hat n^h_{i+1})\rangle}{\langle (1-\hat n^h_i)\rangle} \leq J.
    \label{eq:Jeff}
\end{equation}
This reflects the fact that, in the original model, the Ising interaction acts only when neighboring sites are occupied by particles, whereas in squeezed space it acts uniformly, since holes are absent by construction.
\emph{Detecting hidden order via nonlocal order parameters.--}
As we show in the following, in certain parameter regimes of~\cref{eq:doped_hopping_TFIM}, the interplay between kinetic and magnetic degrees of freedom manifests in spin order being obscured by fluctuating holes, which render the spin order invisible to any local order parameter.
We will highlight this effect at the example of the staggered real space magnetization 
\begin{equation}
    M_S = \langle |\frac{1}{L}\sum_{j=1}^L (-1)^j \hat S^z_j|\rangle\;,
\end{equation}
which suffices as a local order parameter for the conventional (i.e. undoped)~\gls{TFIM}: The magnetization indicates order ($M_S>0$) in the $\mathbb{Z}_2$ symmetry-broken \gls{AFM} phase which occurs at $h<h^c_x=0.5J$. As the transverse field is tuned across the quantum critical point into the~\gls{PM} phase, the magnetization vanishes, see gray triangular datapoints in Fig.~\ref{fig:sqsp} b).

Let us now consider the same observable for the $t_1-$ \gls{TFIM}: The green data points in Fig.~\ref{fig:sqsp} b) show the absence of $M_S$ for any value of $h_x$. In analogy to $M_S$, we now introduce the squeezed space staggered magnetization
\begin{equation}
    M^*_{S} = \langle |\frac{1}{\tilde L}\sum_{\tilde j=1}^{\tilde L}  (-1)^{\tilde j} \hat S^z_{\tilde j}|\rangle_{\rm s}\;,
\end{equation}
which is evaluated on the squeezed space wave function, only. $M^*_{S}>0$ indicates magnetic order in squeezed space for low transverse fields $h_x$, see blue squares in fig~\ref{fig:sqsp} b). The transformation to squeezed space allows us to recover hidden magnetic order which $M_S$ is insensitive to by capturing the holes separately, see Fig.~\ref{fig:sqsp} a). As the field is increased, the model undergoes a \emph{hidden} symmetry-restoring phase transition to the PM phase. This transition is governed by a~\gls{hQCP}  -- i.e. a quantum critical point in squeezed space -- at $h_x \approx 0.3 J$.
Note that both the real space and squeezed space magnetization show a finite-size related offset in the magnetization, which vanishes in the limit of infinite system size. We compare these results to the analytical prediction of the magnetization by Onsager and Yang $M_{OY}(J) = \frac{1}{2}[1-(\frac{2h_x}{J})^2]^{1/8}$, using $J^{\rm snap}_{\rm eff} \approx 0.65 J$ computed via~\cref{eq:Jeff} yielding excellent agreement, see Fig.~\ref{fig:sqsp} $b)$.

The above quantities were computed by performing \acrshort{DMRG} simulations, taking $z$-basis snapshots of the many-body wavefunction via the perfect sampling approach~\cite{ferris_perfect_2012,buser_snapshot-based_2022} and transforming them to squeezed space, all using the \textsc{SyTen} toolkit.
Further details on numerical methods can be found in the Supplementary Material.

\section{Results\label{sec:results}}
\subsection{Phase Diagram\label{sec:phase_diagram}}
\begin{figure}
    \includegraphics[width=.4\textwidth]{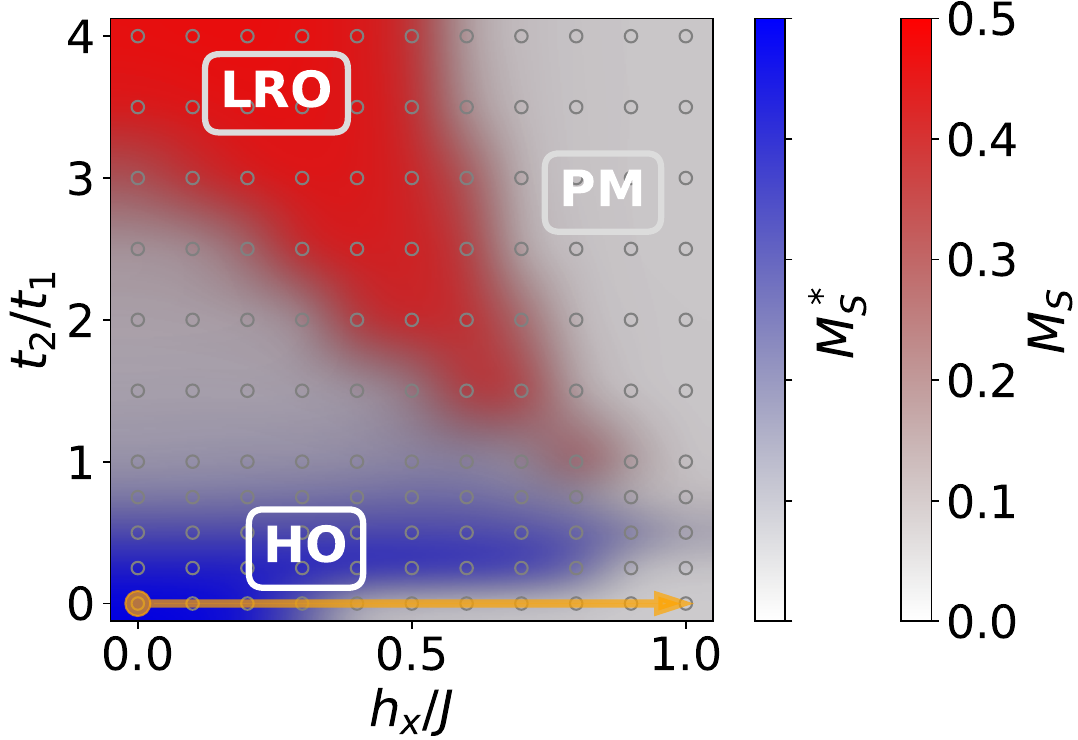}
    \caption{\justifying Numerical study of the phase diagram of the $t_1-t_2-$ \gls{TFIM}~\cref{eq:doped_hopping_TFIM}.
    The figure displays the staggered magnetization in real space, $M_S$ (red), and in squeezed space, $M_S^*$ (blue), represented by two overlaid color maps whose opacity increases with the magnitude of the corresponding order parameter.  
    Extended regions exhibiting \gls{LRO} and \gls{HO} are observed throughout the phase diagram. For intermediate values of $t_2$ and $h_x < 0.5 J$, however, neither type of order is present.  
    The orange arrow indicates the $h_x$ scan shown in Fig.~\ref{fig:sqsp} and Fig.~\ref{fig:dsf}.  
    The phase diagram is constructed from the indicated data points (gray circles), computed for $L=90$, $N_h=24$, and $t_1/ J = 1$.  
    Additional details on the numerical methods are provided in the Supplementary Material.}
    \label{fig:pd}
\end{figure}

We use the observables introduced in the previous section to map out the phase diagram of the $t_1-t_2$-\gls{TFIM}. Fig.~\ref{fig:pd} shows the phase diagram as a function of the~\gls{NNN} hopping amplitude $t_2$ and the transverse field $h_x$. It consists of two layered heatmaps indicating the real (squeezed) space magnetization  $M_S$ ($M^*_S$) in red (blue). The opacity is proportional to the absolute value of the respective order parameter.
Note that the orange arrow in Fig.~\ref{fig:pd} indicates the $h_x$ scan displayed in Fig.~\ref{fig:sqsp}, Fig.~\ref{fig:dsf} and Fig.~\ref{fig:dsf_x}. The phase diagram features the following phases:

(i) Hidden order phase (\gls{HOP}): $M^*_S > 0, M_S = 0$. In this phase, the spins exhibit hidden \gls{SSB}: The antiferromagnetic order is obscured by holes, leading to a vanishing real space magnetization $M_S = 0$. The squeezed space magnetization captures the hidden~\gls{LRO}, $M^*_S > 0$. The \gls{HOP} extends to finite values of $t_2 < t_1$.

(ii) Long-range ordered phase (\gls{LRO}): $M^*_S = 0, M_S > 0$. This phase is ordered in accordance with the Landau-Ginzburg paradigm: Spontaneous breaking of the $\mathbb{Z}_2$-spin symmetry is associated with \gls{LRO} indicated by a finite value of the local order parameter $M_S$. This phase occurs for $t_2 \gg t_1$ where the terms that preserve sublattice parity dominate.

(iii) Paramagnetic phase (PM): $M^*_S = 0, M_S = 0$.
The \gls{PM} phase is characterized by the vanishing real and squeezed space order parameter, indicating the absence of both conventional and hidden order. This phase is reached by increasing the transverse field $h_x$ across the (hidden) \gls{QCP} starting from the (hidden) ordered phase.
%
%
In fact, it can be shown that a $t_2$ scan starting from the~\gls{HOP} undergoes two $k_F$-changing transitions before reaching the~\gls{LRO} phase. However, investigating this regime lies outside of the scope of the current work. 

Instead, it is the goal of this paper to investigate the spectral features of the $t_1-t_2-$~\gls{TFIM}. While these are well understood for the~\gls{LRO} and the \gls{PM} phase, in this paper we focus on the spectral signatures of hidden order. In particular, we investigate how they emerge from the composition of spectral features of well-known charge and spin models.

\subsection{Numerical Approach\label{sec:numerical_sf}}
\begin{figure*}
	\includegraphics[width=\textwidth]{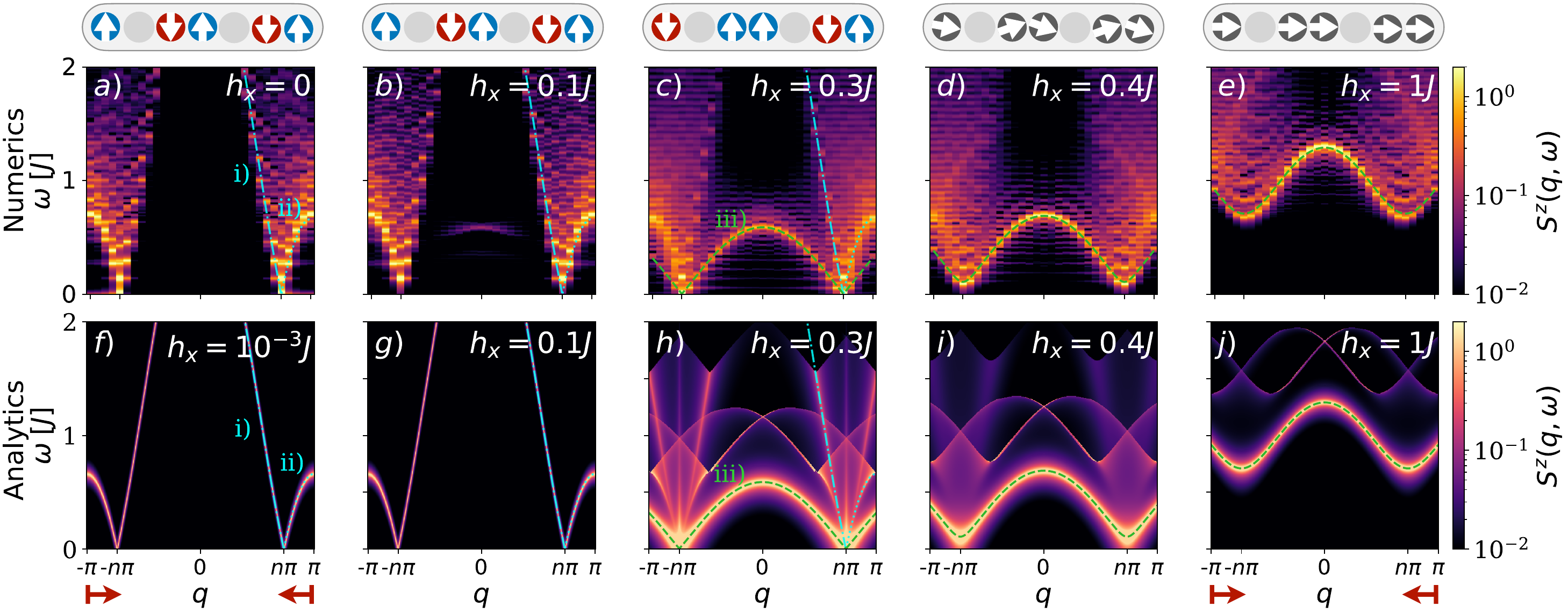}
	\caption{\justifying
		Comparison between numerical and analytical results for the dynamical spin structure factor $S^{z}(q,\omega)$ of the $t_1$-\gls{TFIM}~\cref{eq:doped_hopping_TFIM} for an $h_x$ scan across the hidden~\gls{QCP} at $h^c_x = 0.3J$.
        The schematics at the top of the figure illustrate the corresponding ground-state configurations on which the excitations are performed. 
        $a)$--$e)$ Numerically computed \gls{DSF} $S^{z}(q,\omega)$ across the transition from the~\gls{HO} phase to the~\gls{PM} phase.  
        $f)$--$j)$ Analytical results for the corresponding \gls{DSF}, obtained from the expressions $S^{z}_{\rm O}(q,\omega)$ for $h_x< h_x^c$ [$f)$--$g)$] and $S^{z}_{\rm PM}(q,\omega)$ for $h_x\geq h_x^c$ [$h)$--$j)$].  
        For $h_x< h_x^c$, we highlight the linear dispersive branches [feature $i)$], corresponding to eq.~\eqref{eq:dsf_ordered} by dash-dotted cyan lines. Towards the edge of the Brillouin zone the linear dispersion experiences lattice effects [feature $ii)$, dotted cyan line].
        For $h_x\geq h_x^c$, we further indicate the rescaled Bogoliubov dispersion [feature $iii)$] corresponding to the first term in eq.~\eqref{eq:dsf_disordered} by a green dashed line.
        The red arrows indicate the shift of the gapless points (or spectral minima) from $q=\pi$ in the conventional~\gls{TFIM} to $q=n\pi$ in the $t_1$-\gls{TFIM}.  
        The numerical results are shown for a system with $L=30$, $N_h=8$, and $t_1 / J = 1$.  
        Additional details on the implementation are provided in the Supplementary Material.
	\label{fig:dsf}}
\end{figure*}
\begin{figure*}
	\includegraphics[width=\textwidth]{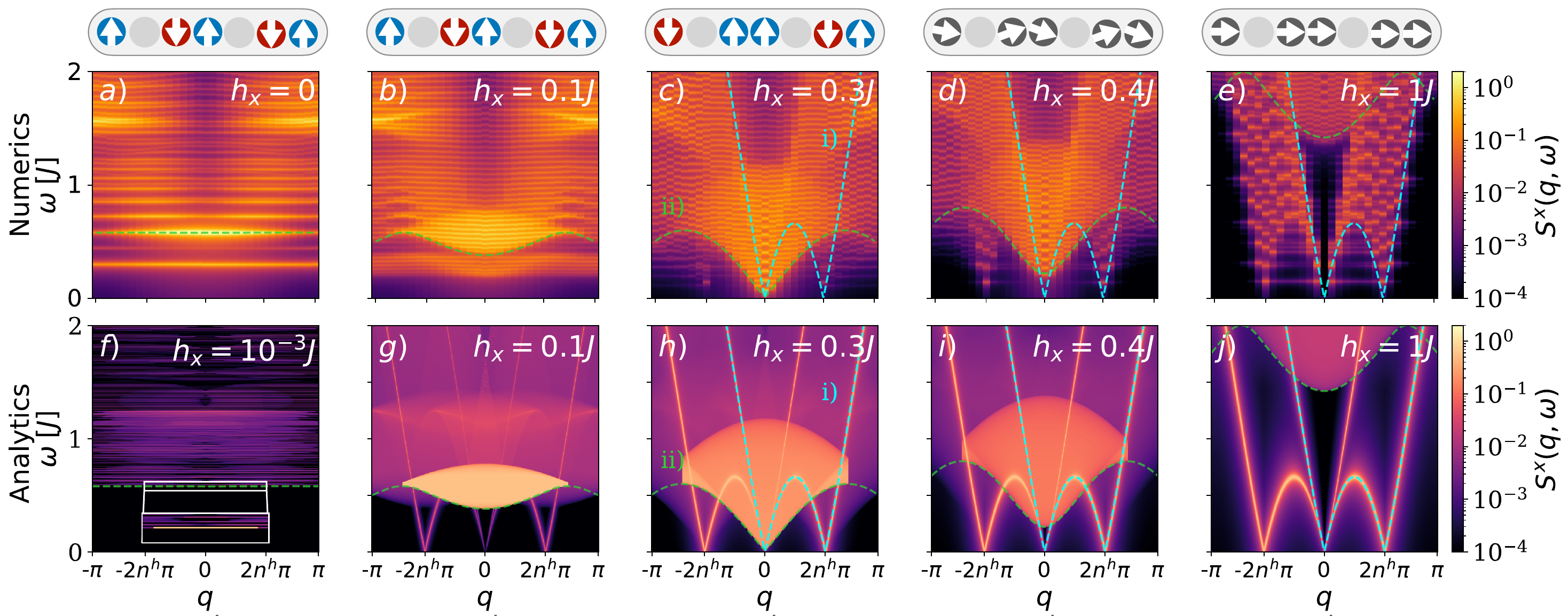}
    \caption{\justifying
    Comparison between numerical and analytical results for the dynamical spin structure factor $S^{x}(q,\omega)$ of the $t_1$-\gls{TFIM}~\cref{eq:doped_hopping_TFIM} for an $h_x$ scan across the hidden \gls{QCP} at $h^c_x = 0.3J$.
    The schematics at the top of the figure illustrate the corresponding ground-state configurations on which the excitations are performed.  
    $a)$--$e)$ Numerically computed \gls{DSF} $S^{x}(q,\omega)$ across the transition from the~\gls{HO} phase to the~\gls{PM} phase.  
    $f)$--$j)$ Analytical results for the corresponding \gls{DSF} across the same transition.
    The inset in $f)$ zooms in on the lowest-lying analytical energy level. 
    Throughout all subplots, the analytical lower bound of the spin-continuum in eq.~\eqref{eq:sx_dsf} is indicated by a dashed green line [feature $ii)$].
    We further highlight the linear dispersive branch in eq.~\eqref{eq:sx_dsf} by a dashed cyan line [feature $i)$].
    The numerical results are shown for a system with $L=30$, $N_h=8$, and $t_1 / J = 1$. 
\label{fig:dsf_x}}
\end{figure*}

We investigate the longitudinal ($\alpha = z$, Fig.~\ref{fig:dsf}) and transverse ($\alpha = x$  Fig.~\ref{fig:dsf_x})~\gls{DSF} of the $t_1$-\gls{TFIM} for an $h_x$ scan across the hQCP, separating the~\gls{HO} from the SRO phase and indicated by the orange arrow in Fig.~\ref{fig:pd}.
The~\gls{DSF} is evaluated numerically following the definition
\begin{equation}
\begin{aligned}
		S^{\alpha}(q,\omega) & = \frac{1}{L} \sum_{i= 1}^L  \mathrm{e}^{-iqr_i} \int^{\infty}_{-\infty} dt e^{i\omega t} \langle \hat S^\alpha_i(t) \hat S^\alpha_0(0)\rangle_{cc}.
		\label{eq:dsf}
\end{aligned}
\end{equation}
Here, the time-dependent connected correlator with respect to the ground state wave function $\ket{\psi_0}$ is given by $\langle \hat S^\alpha_i(t) \hat S^\alpha_0(0)\rangle_{cc} = \bra{\psi_0} \hat U^\dagger(t) \hat S^\alpha_i\hat U(t) \hat S^\alpha_0\ket{\psi_0} - \bra{\psi_0}\hat S^\alpha_i\ket{\psi_0}\bra{\psi_0}\hat S^\alpha_0\ket{\psi_0}$ with the time evolution operator $\hat U(t) = e^{-it\hat H/\hbar}$.
The ground states $\ket{\psi_0}$ were obtained performing~\gls{DMRG} ground-state searches with a maximum bond-dimension of the underlying~\gls{MPS} representations $\chi_\mathrm{max} = 3000$~\cite{white_density-matrix_1993,schollwock_density-matrix_2005,schollwock_density-matrix_2011}.
For increased resolution, instead of computing~\cref{eq:dsf} directly from a vanilla real-time evolution and subsequent Fourier transformation on a finite time-domain, we employed the recently introduced~\gls{CTKS} expansion~\cite{paeckel_spectral_2026}. For more detailed information on the numerical approach we refer to the Supplementary Material.
Our results for the  $z$-\gls{DSF} are displayed in the upper row of Fig.~\ref{fig:dsf}.
At $h_x = 0$, see Fig.~\ref{fig:dsf} a), two gapless points at $q=\pm n\pi$ are present, with the density $n=N/L$.
Originating from these gapless points, a dispersive branch emerges: While at the~\gls{BZ} center this branch follows a linear behavior (feature $i)$), lattice effects become visible close to the~\gls{BZ} edge and cause the curving of the branch (feature $ii)$).
At $h_x = 0.1 J$, see Fig.~\ref{fig:dsf} b), a weak spectral signal emerges at the~\gls{BZ} center which gains in spectral weight as the field is increased further: At the critical field $h_x = 0.3 J$, see Fig.~\ref{fig:dsf} c), a cosine dispersion with minima at $q=\pm n\pi$ has appeared (feature $iii)$). The structure of the remaining signal is left unchanged.
When increasing the transverse field strength further, the spectral weight shifts away from the gapless points and towards the~\gls{BZ} center.
Simultaneously, a gap opens which increases as $h_x$ is increased further, see Fig.~\ref{fig:dsf} d-e).

From comparison with the~\gls{DSF} of the (undoped)~\gls{TFIM}, see Supplementary Material, we find that the presence of charges introduces a doping dependent momentum shift of the spectral signatures: The gapless momenta $q = \pm \pi$ -- indicating \gls{AFM} order in the \gls{TFIM} -- are shifted to $q=\pm \pi n$ in the~\gls{HOP}. Likewise, the Bogoliubov dispersion, characterizing the excitation spectrum of the \gls{TFIM} in the \gls{PM} phase, is rescaled, now yielding minima at $q =\pi n$ instead of $q=\pi$.

In the upper row of Fig.~\ref{fig:dsf_x} we display the results for the $x$-\gls{DSF}.
At zero transverse field, see Fig.~\ref{fig:dsf_x} a), the dynamical structure factor exhibits flat bands, one of which carries a pronounced spectral weight. As the transverse field is increased to $h_x =0.1 J$, see Fig.~\ref{fig:dsf_x} b), the pronounced 
feature broadens into a continuum. Around the critical field $h_x=0.3 J$, the continuum becomes gapless, see feature $ii)$ in Fig.~\ref{fig:dsf_x} c), and two additional gapless points appear at $2k_F^c = 2\pi n^h$ with $n^h= 1-n$, see feature $i)$. Increasing the transverse field further causes the continuum to gap out while the gapless features at $q = 0, \pm 2k_F^c$ remain. 

Recall that in the undoped~\gls{TFIM} at $h_x=0$, the~\gls{DSF} is dominated by a flat band at $\omega = J$, see Supplementary Material. In comparison, doping the TFIM causes a splitting of the single flat band into several finer bands.  In the disordered phase, we find that doping renormalizes the two quasi-particle continuum with respect to the~\gls{TFIM} and an additional charge and continuous background signal appears.

Understanding the renormalization and origin of the spectral features motivates the analytical treatment of the~\gls{DSF} in the next section. We will then return to a combined interpretation of the numerical and analytical results.

\subsection{Analytical Approach\label{sec:analytical_sf}}
In order to analytically compute the~\gls{DSF} defined in~\cref{eq:dsf}, we utilize the concept of \textit{squeezed space}, introduced in Sec.~\ref{sec:model}.
We therefore assume a factorization of spin and charge degrees of freedom, which is exact in the $t_1 \rightarrow \infty$ limit of the $t_1-$~\gls{TFIM}.
We make a corresponding ansatz for the Hamiltonian~\cref{eq:doped_hopping_TFIM}, $\hat H = \hat H_{s} \otimes \mathds{1}_c + \mathds{1}_s \otimes \hat H_{c}$:
The holes $\hat h^{(\dagger)}_i$ are subject to a spinless free fermion Hamiltonian of length $L$, while the spin \gls{DOF} are described by a \gls{TFIM} of length $\tilde{L} = L-N^h$,
\begin{align}
        \hat H_c &= -t \sum_{\langle ij\rangle} (\hat h^\dagger_i \hat h_j + \mathrm{h.c}),\label{eq:Hc}\\
        \hat H_s &= \tilde J  \sum_{\langle \tilde{i}\tilde{j}\rangle} \hat \sigma^z_{\tilde{i}}  \hat \sigma^z_{\tilde{j}} + \tilde h  \sum_{\tilde{j}} \hat\sigma^x_{\tilde{j}}\;.\label{eq:Hs}
\end{align}
For the sake of clarity we denote the spin \gls{DOF} on squeezed space site $\tilde{i}$ in terms of the Pauli-$\alpha$ operators $\hat \sigma^\alpha_{\tilde{i}}$. The amplitudes are related to those of the original model, eq.~\eqref{eq:doped_hopping_TFIM}, via $\tilde{J}=J_{\rm eff}/4$, $\tilde{h} =h_x/2$ and $t=t_1$. Note that the use of $J_{\rm eff}(n)$ takes into account the coupling of spin and charge on a mean-field level.

$\hat H_s$ can be solved exactly via a Bogoliubov transformation yielding the dispersion 
\begin{equation}
    \varepsilon^s_k = 2\sqrt{\tilde{J}^2 + \tilde{h}^2 + 2\tilde{J}\tilde{h} \cos (ak)}. 
    \label{eq:bogo}
\end{equation}
A gap closing in $\varepsilon^s_k$ indicates the critical point of the \gls{TFIM} at $\tilde{h} = \tilde{J}$. 
Likewise, $\hat H_c$ is exactly solvable in momentum space with single-electron dispersion $\varepsilon^c_k = -2t\cos(ak)$.

To evaluate the expectation value~\cref{eq:dsf} in squeezed space, we explicitly account for the charge degree of freedom $ \hat S^\alpha_i \rightarrow \hat S^\alpha_i \hat n_i$ in order to reduce the local Hilbert space dimension from three ($S^\alpha_i = \pm\frac{1}{2},h$) to two ($S^\alpha_i =  \pm\frac{1}{2}$).
The expectation value then decouples into
\begin{equation}
\begin{aligned}
    \langle \hat S^\alpha_i (t) \hat S^\alpha_0(0)\rangle = \langle \hat S^\alpha_{i}(t) \hat n_{i}(t) \hat S^\alpha_{0}\hat n_{0}\rangle \approx \langle \hat S^\alpha_{\tilde i}(t) \hat S^\alpha_{0}\rangle_s  \langle \hat n_{i}(t) \hat n_{0}\rangle_c\;. \notag
\end{aligned}
\end{equation}
The assumption of spin charge separation within the squeezed space construction has two main implications when computing the dynamical structure factor analytically: The ansatz for the Hamiltonian,~\cref{eq:Hc} and ~\cref{eq:Hs}, as well as the factorization of the expectation value.
We furthermore make the assumption that real and squeezed space sites are related via $\tilde{i} = in$, using the mean density $n$.
In order to sum over these rescaled coordinates, we introduce the continuum variable $x = ai$ so that $\tilde{i} \rightarrow xn/a$. Consequently, the sum is replaced by an integral $\sum_{j} \rightarrow \frac{1}{a} \int dx$, and the spin and charge~\gls{DOF} become continuous, $\hat n_j \rightarrow\rho(x)$ and $\hat S_j \rightarrow \sigma(x)/2$.
In the continuum limit of squeezed space, the \gls{DSF} is then given by
\begin{widetext}
\begin{equation}
\begin{aligned}
    S^{\alpha}_P(q,\omega) &\approx\int dt \int dx \;e^{-i(qx - \omega t)} [\langle \hat \sigma^\alpha(xn,t) \hat \sigma^\alpha(0,0)\rangle_s  \langle \hat \rho(x,t) \hat \rho(0,0)\rangle_c - 
    \langle\hat \sigma^\alpha(xn,0)\rangle_s \langle \hat \sigma^\alpha(0,0)\rangle_s \langle \hat \rho(x,0)\rangle_c \langle\hat \rho(0,0)\rangle_c]\\
    &= c \int dt \int dk \;  e^{iwt} [ S^{\alpha, \rm TFIM}_{\rm P}(k/n,t) S_\rho(q-k,t) - S^{\alpha, \rm TFIM}_{\rm P,\rm stat}(k/n) S_{\rho,\rm stat}(q-k) ]\;.
    \label{eq:sf_squeezed}
\end{aligned}
\end{equation}
\end{widetext}
In the last equality we have exploited the convolution theorem, $\mathcal{F} \{ f(x)\cdot g(x)\} = (F*G)(k)$ with the Fourier transform $F(k) = \mathcal{F} \{ f(x)\}=\int_{-\infty}^{\infty} dx \, e^{-iqx} f(x)$.
The convolution of two functions is defined as $(F*G)(k) = \int_{-\infty}^{\infty} dq\; F(q)G(k-q)$.
Furthermore, P denotes the ordered (O) and paramagnetic (PM) phase of the \gls{TFIM}, which require different analytical approaches for $\alpha = z$. $c$ denotes a constant specified in the Supplementary Material.

According to~\cref{eq:sf_squeezed}, in the continuum limit of squeezed space, the dynamical structure factor~\cref{eq:dsf} can be expressed as a momentum convolution of the individual spin and charge contributions.
The above expression clearly highlights the effect of the squeezed space construction:
The real space scaling $x \rightarrow x n$ leads to an inverse scaling of the momenta $q \rightarrow q/n$ in the spin contributions, only.
We conclude that the $\alpha$-\gls{DSF} of the doped \gls{TFIM}~\cref{eq:doped_hopping_TFIM} can be approximated as a convolution of the rescaled $\alpha$-\gls{DSF} of the \gls{TFIM} and a charge structure factor of the spinless free fermion model,~\cref{eq:Hc} and ~\cref{eq:Hs}.
Evaluating~\cref{eq:sf_squeezed} requires the expressions of the~\gls{DSF} for the individual models and $\alpha$. A summary of the required terms can be found in~\cref{tab:spin_models}.

\emph{Structure Factors of Individual Models.--}
In the following, we state the dynamical structure factors of the individual models, that are required to compute~\cref{eq:sf_squeezed}.
For a derivation of these textbook results we refer to the Supplementary Material.

\textbf{Free fermions:} The charge structure factor $S_\rho(q,t)$ can be derived from the fermionic single-particle Green's function and is given by
\begin{equation}
\begin{aligned}
S_\rho(q,t)\simeq
\underbrace{2\pi \rho_0^{\,2}\,\delta(q)}_{S^0_\rho(q)}
+
\underbrace{\frac{|q|}{2\pi}\,e^{-i |q| v^c_F t}}_{S^1_\rho(q,t)}\;,
\label{eq:FT_charge_corr}
\end{aligned}
\end{equation}
with the charge Fermi velocity $v^c_F = \partial_k \varepsilon^c_k|_{k=k^c_F}$. 
The charge Fermi momentum is given by $k^c_F = n^h\pi$.
$S^0_\rho(q)$ corresponds to the static background density and $S^1_\rho(q)$ describes the propagation of long-wavelength density waves.
We neglect the $2k_F$-oscillatory Friedel term as it does not contribute dominant spectral features in our analysis.
The static structure factor is given by $S_{\rho,\rm stat}(q) =S^0_\rho(q)$.\\

\textbf{Disordered TFIM, longitudinal DSF:} The dynamical $z$-structure factor can be obtained via a form factor expansion yielding
\begin{equation}
\begin{aligned}
S_{\rm PM}^{z,\rm{TFIM}}(q, t) = \frac{\bar s^2}{2 \Omega^{s,\pm}_{q}}\cos(\Omega^{s, \pm}_{q} t)\;,
 \label{eq:FT_spin_disordered_corr}
\end{aligned}
\end{equation}
with the Bogoliubov dispersion expanded around the minima at $q = \pm \pi/a$, $\Omega^{s,\pm}_{q}=(m^2 + (v^s_F)^2 (q\pm \pi/a)^2)^{1/2}$.
The effective velocity is given by $v^{s}_{F} = 2a\sqrt{\tilde{J}\tilde{h}}$ and we identify the mass gap $m = 2|\tilde{J} - \tilde{h}|$.
$\bar s^2= \abs{F_1}^2 \propto m^{1/4}$ denotes the absolute square of the one-particle form factor~\cite{wu_real-time_2026}.
Due to $\langle \hat \sigma^z\rangle=0$ in the \gls{PM} phase, the static term $S^{z,\rm{TFIM}}_{PM,\rm stat}(k)= 0$ vanishes.\\

\textbf{Ordered TFIM, longitudinal DSF:} We approximate the \gls{DSF} in the ordered phase, in which spin excitations are largely suppressed, as 
\begin{equation}
   S^{z,\rm{TFIM}}_{\rm O} (q,t) = \pi \tilde{M}_{OY}^2 \delta\left(q\pm \frac{\pi}{a}\right)\;,
    \label{eq:FT_spin_ordered_corr}
\end{equation}
with the Onsager-Yang result for the magnetization $\tilde{M}_{OY} = [1-(\frac{\tilde{h}}{\tilde{J}})^2]^{1/8}$.
The static term gives the same contribution, i.e. $S^{z,\rm{TFIM}}_{O,\rm stat}(q)=S^{z,\rm{TFIM}}_{\rm O}(q)$.\\

\textbf{TFIM, transverse DSF:}
The $x$-\gls{DSF} is obtained by relating the transverse spin operator $\sigma^x$ to the energy density operator, see Supplementary Material. The full dynamical structure factor is defined as $S^{x,\rm{TFIM}}(q,t) = S^{x,\rm{TFIM}}_{\rm{conn}}(q,t) +  S^{x,\rm{TFIM}}_{\rm{stat}}(q)$, with the connected part
\begin{equation}
\begin{aligned}
S^{x,\rm{TFIM}}_{\rm{conn}}(q,t) = \int_{-\infty}^{\infty} &\frac{dk}{4\pi} e^{-i (\Omega^s_k + \Omega^{s}_{q-k}) t} \\
&\cdot \left( 1 - \frac{m^2 + (v^s_F)^2 k (q-k)}{\Omega^s_k \Omega^s_{q-k}} \right) \;,
\label{eq:FT_spin_x}
\end{aligned}
\end{equation}
and the expanded Bogoliubov dispersion $\Omega^{s}_{q}=(m^2 + (v^s_F q)^2)^{1/2}$.
The static structure factor is given by $S^{x,\rm{TFIM}}_{\rm stat}(q)=\langle \sigma^x \rangle^2\delta(q)$.

\subsubsection{Longitudinal dynamical structure factor}
\begin{table}[h]
    \centering
    \renewcommand{\arraystretch}{1.8}
    \begin{tabular}{c | c || c | c | c}
         \makecell{$t_1-$TFIM\\ DSF $S^{\alpha}_P$} & \makecell{TFIM DSF\\ $S^{\alpha, \rm TFIM}_{\rm P}$} & charge & spin & \makecell{composite } \\
        \hline
        $S^z_{\rm HO}$,~\cref{eq:dsf_ordered} & $S^{z,\rm{TFIM}}_{\rm O}$,~\cref{eq:FT_spin_ordered_corr} & $\cross$ & & \\
        \hline
        $S^z_{\rm PM}$,~\cref{eq:dsf_disordered} & $S^{z,\rm{TFIM}}_{\rm PM}$,~\cref{eq:FT_spin_disordered_corr} & & $\cross$ & $\cross$ \\
        \hline
        $S^x$,~\cref{eq:sx_dsf}      & $S^{x,\rm{TFIM}}$,~\cref{eq:FT_spin_x}     & $\cross$ & $\cross$ & $\cross$ \\
    \end{tabular}
     \caption{Overview of the analytical approach. In order to evaluate the~\gls{DSF} of the $t_1-$TFIM, $S^{\alpha}_P$ (first column from the left), we use the expression derived in~\cref{eq:sf_squeezed}. The first term requires the charge structure factor $S_\rho(q,t)$,~\cref{eq:FT_charge_corr}, and the respective expression for spin structure factor $S^{\alpha, \rm TFIM}_{\rm P}$ (second column from the left). The second term of~\cref{eq:sf_squeezed} requires their static counterparts. The three columns to the right indicate the kind of terms that the convolution yields: Renormalized spin or charge excitations, or composite spin-charge 
excitations. For $\alpha = z$ the analytical approach requires a distinction between the ordered (O) and paramagnetic (PM)~\gls{TFIM} phase, yielding the hidden order (HO) and PM phase of the $t_1-$TFIM, respectively.}
    \label{tab:spin_models}
\end{table}

The dynamical structure factor for the~\gls{HO} phase is obtained as the momentum convolution, see~\cref{eq:sf_squeezed}, of the charge structure factor~\cref{eq:FT_charge_corr} and the spin-$z$ structure factor~\cref{eq:FT_spin_ordered_corr}  in the ordered phase. Performing the convolution yields
\begin{equation}
S^{z}_{\rm HO}(q,\omega) = \pi \tilde{M}_{OY}^2 |q\pm Q_n| \delta(\omega - v^c_F |q\pm Q_n|)\;.
\label{eq:dsf_ordered}
\end{equation}
For intermediate steps we refer to the Supplementary Material. Since the spin degrees of freedom are effectively frozen in the spin-ordered phase, the dynamical structure factor is dominated by the direct excitation of acoustic charge modes governed by $v^c_F$. The gapless points are located at the doping-dependent magnetic ordering vector $Q_n = \frac{\pi n}{a}$, see Fig.~\ref{fig:dsf} $f), g)$. The spectral weight is governed by the magnetization $M_{OY}$ and the distance from the gapless points.
In the \gls{PM} phase, the \gls{DSF} is obtained as the momentum convolution of the charge structure factor eq.~\eqref{eq:FT_charge_corr} and the spin-$z$ structure factor in the disordered phase eq.~\eqref{eq:FT_spin_disordered_corr} and takes the following form
\begin{equation}
\begin{aligned}
& S^{z}_{\rm PM}(q,\omega) = \frac{\pi^2 \rho_0^2 \bar s^2}{\Omega^{s,\pm}_{q/n}} \delta(\omega - \Omega^{s,\pm}_{q/n}) \\ 
&+ \frac{\bar s^2}{4}\int dk \frac{|k|}{\Omega^{s,\pm}_{(q-k)/n}} \delta(\omega - \Omega^{s,\pm}_{(q-k)/n} - |k|v_F^c)\;,
\label{eq:dsf_disordered}
\end{aligned}
\end{equation}
The first term corresponds to a probe creating a single spin-wave that obeys the rescaled Bogoliubov dispersion (see feature $iii)$ in Fig.~\ref{fig:dsf} $h)$) without exciting the charge sector. The second term describes a composite excitation, i.e. the simultaneous creation of a spin and a charge excitation sharing the total momentum $q$ and energy $\omega$. This interplay introduces a threshold condition for the energetic cost of excitations $\omega \geq m + \abs{k} v^c_F$. Below this threshold, no excitations exist while above, a continuum of excitations opens. Note that $m$ is the cost of a magnetic excitation (cf. \gls{TFIM}) and the charge sector can take arbitrary momentum at an energetic cost $\propto \abs{k}$. In Fig.~\ref{fig:dsf}, these excitations correspond to the continuum in $h)-j)$.

A few comments are in order regarding the evaluation of the above expressions: The continuum field theory is blind to the periodic lattice and assumes the~\gls{BZ} boundary is at infinity. To enable a more direct comparison with the lattice setting realized in experiments and in our numerical simulations, we account for both the lattice dispersion and the momentum dependence of the Fermi velocity in the analytical treatment. This approach allows us to analytically capture the spectral weight beyond the lowest order signal obtained from the continuum description. The full continuum results can be found in the Supplementary Material. In order to avoid back-scattering at the~\gls{BZ} boundaries that would occur from the periodic dispersion, but is not native to the continuum description, we project out back-scattering momenta. This process leads to numerical artifacts which manifest in the spin-charge continuum~\cref{eq:dsf_disordered}. The integral in~\cref{eq:dsf_disordered} is evaluated using the property of the Dirac delta function $\int f(k) \delta(g(k)) dk = \sum_{i} \frac{f(k_i)}{|g'(k_i)|}$ where $k_i$ are the roots (i.e. kinetically allowed momenta) such that $g(k_i) = 0$.

\subsubsection{Transverse dynamical structure factor}
The $x$-\gls{DSF} is given as the momentum convolution (see~\cref{eq:sf_squeezed}) of the charge structure factor~\cref{eq:FT_charge_corr} and the spin-$x$ structure factor~\cref{eq:FT_spin_x} which yields
\begin{equation}
\begin{aligned}
&S^x(q,\omega) = \rho_0^2 S^x_{\rm{conn}}\left(\frac{q}{n}, \omega\right) \\& + \int \frac{dk}{(2\pi)^2} \abs{k} S^x_{\rm{conn}}\left( \frac{q - k}{n}, \omega 
- |k|v^c_F \right) \\ &+ \frac{\langle \sigma^x \rangle^2 |q|}{2\pi} \delta(\omega - |q|v^c_F)\;,
\end{aligned}
\label{eq:sx_dsf}
\end{equation}
with $\langle \hat{\sigma}^x \rangle$ computed in the Supplementary Material.
The first term describes a process where the dynamical probe exclusively excites the massive two-particle Ising continuum, while the charge field remains strictly in its ground state, providing only a static background density $\rho_0^{\,2}$. The kinematic threshold for this term is identical to the pure rescaled Ising model: $\omega \geq \sqrt{(v^s_F)^2 (q/n)^2 + 4m^2}$, see Fig.~\ref{fig:dsf_x} feature $ii)$. 
The second contribution captures a composite excitation of a gapless charge mode and two spin excitations. The total momentum $q$ and energy $\omega$ must be partitioned, i.e. an amount of momentum $k$ is absorbed by the charge field to create a gapless excitation with energy $|k|v^c_F$. The remaining momentum $q - k$ and remaining energy $\omega - |k|v^c_F$ must subsequently excite the massive two-particle continuum of the Ising spins. The absolute minimum energy required to observe a signal here is $\omega_{th}(q) = \min_{k} \left[ |k|v^c_F + \sqrt{(v^s_F)^2 \left((q-k)/n\right)^2 + 4m^2} \right]$.
The last term in~\cref{eq:sx_dsf} captures the frozen spin sector. Consequently, the external probe transfers all of its momentum $q$ and energy $\omega$ exclusively to a single acoustic density fluctuation. The observed signal is sharp and follows the pure acoustic dispersion relation precisely at $\omega = |q|v^c_F$ and increases in intensity as $\tilde h$ is increased, see Fig.~\ref{fig:dsf_x} feature $i)$.

\section{Effect of Doping on the longitudinal DSF\label{sec:HO_discussion}}

Combining our numerical and analytical results, we now discuss the effect of doping on the excitation spectrum, focusing on the longitudinal~\gls{DSF} shown in Fig.~\ref{fig:dsf}. Both approaches consistently demonstrate that hole doping induces a doping-dependent renormalization of the spectral features of the \gls{TFIM}, corresponding to the momentum rescaling $q \rightarrow q/n$, see~\cref{eq:dsf_ordered}. Note that a similar effect has been observed experimentally by Li et al.~\cite{li_doping_2025} who studied the doping dependence of~\gls{RIXS} spectra of 1D cuprate materials. In particular, the gapless points at $q=\pm\pi$, which signal \gls{AFM} order in the undoped \gls{TFIM}, are shifted to $q=\pm n\pi$ in the~\gls{HOP}. This reflects the elongation of the wavelength of spin excitations caused by the presence of holes, although for the incommensurate fillings considered here this does not directly imply an enlarged magnetic unit cell. The analytical expression in~\cref{eq:dsf_ordered} further shows that the dispersive spectral features of the~\gls{HO} phase [features $i)$ and $ii)$ in Fig.~\ref{fig:dsf}a),f)] originate from free fermions propagating through the ordered spin background. This interpretation is reinforced by the relation between the charge and spin structure factors in the~\gls{HOP},
\begin{equation}
S^{z}_{\rm HO}(q,\omega) = S_\rho(q-n\pi,\omega)\;,
\end{equation}
which demonstrates that the gapless spectral response in the hidden-order regime ($h_x<0.3 J$) can be entirely attributed to hole excitations moving coherently through the ordered background. A derivation can be found in the Supplementary Material.

The hidden quantum phase transition between the~\gls{HO} and~\gls{PM} phases is consistently marked by the opening of a spectral gap at a critical field set by the effective Ising interaction $J_{\rm{eff}}$.

In the disordered phase, the numerical and analytical results also show good qualitative agreement. Equation~\cref{eq:dsf_disordered} predicts a renormalized Bogoliubov dispersion with minima at $q=\pm n\pi$, clearly visible in both the numerical and analytical spectra [feature $iii)$ in Fig.~\ref{fig:dsf}c),h)]. In addition, the analytical theory predicts a broad continuum arising from composite spin-charge excitations, which retains characteristic features of the free-fermion response and is likewise observed in the numerical results for $h_x \geq 0.3 J$.

\section{Spectral Signatures of Hidden Order in the transverse DSF\label{sec:spectral_signatures_discussion}}
In this section, we specifically focus on the spectral signatures of hidden order through the transverse dynamical structure factor, shown in Fig.~\ref{fig:dsf_x}. Our analytical approach, eq.~\eqref{eq:sx_dsf}, identifies three distinct contributions (see~\cref{tab:spin_models}) that can be directly associated with features in the numerical spectra. In the hidden-order regime, i.e. for small transverse fields, the pure charge contribution is strongly suppressed, while the dominant response arises from spin and composite spin-charge excitations. Based on this correspondence, we propose experimentally accessible spectral signatures of the hidden-order phase and clarify their microscopic origin.

To this end, recall that in the undoped~\gls{TFIM} at $h_x=0$, the~\gls{DSF} is dominated by a flat band at $\omega=J$ (see Supplementary Material). This feature corresponds to the creation of two dispersionless domain walls upon acting with $\hat{\sigma}^x$ on the ordered spin background, with an energy cost determined by the Ising interaction. As the transverse field increases, the domain walls acquire a dispersive character and the flat band broadens into a continuum.
In the doped model, we identify the pronounced feature at $\omega \approx 0.58 J$, which continuously broadens into the excitation continuum as the quasiparticle dispersion becomes nontrivial, with the renormalized two-quasiparticle excitation energy. From this feature, we extract an effective interaction strength of $J^s_{\mathrm{eff}} \approx 0.58$ which is in good agreement with the prediction from snapshots $J^{\rm snap}_{\mathrm{eff}} \approx 0.65$, see~\cref{eq:Jeff}. This confirms the expected location of the hidden quantum critical point at $h_x/J_{\rm eff} \approx 0.5$.

In addition, we associate the spectral feature at $\omega \approx 0.3J$ with the creation of a single domain wall. In contrast to the undoped case, the presence of holes allows $\hat{\sigma}^x$ to violate only a single Ising bond even within the ordered background. These processes split the single flat band of the undoped~\gls{TFIM} into several finer spectral branches.
Taken together, the defining characteristic of the hidden-order phase is a renormalized and broadened two-particle continuum embedded within a continuous spin-charge background. Outside the hidden-order regime, the dominant spectral features of the doped~\gls{TFIM} can largely be understood as renormalized counterparts of the excitations already present in the undoped model.

\section{Discussion}

In this work, we demonstrated that hidden-order phases leave clear and experimentally accessible fingerprints in dynamical response functions, despite remaining invisible to conventional local equal-time correlators.
Using a combination of matrix-product-state simulations and analytical calculations, we investigated the dynamical structure factor of the doped transverse-field Ising model, the $t_1$-\gls{TFIM}, a minimal system in which charge fluctuations give rise to a hidden-order phase characterized by hidden spontaneous symmetry breaking and the absence of conventional long-range order.

With analytical calculations we showed that the dynamical structure factor of the full model can be expressed as a convolution of the spin and charge sectors. This decomposition allows us to directly connect the numerically obtained spectral response to microscopic processes and provides a transparent physical interpretation of the excitation spectrum across the hidden-order and paramagnetic regimes, albeit being in a strongly-correlated doped model.

A central result of our work is that hole doping induces a universal momentum renormalization of the spectral features, $q \rightarrow q/n$, relative to the undoped~\gls{TFIM}. In the hidden-order phase, this leads to a shift of the gapless points from $q=\pm\pi$ to $q=\pm n\pi$: Doping distorts the underlying antiferromagnetic spin pattern, effectively increasing the dominant wavelength of spin excitations. The resulting low-energy response is dominated by coherent hole propagation through the ordered spin background, producing dispersive free-fermion-like excitations. 
In the transverse dynamical structure factor, hidden order is characterized by a renormalized and broadened two-particle continuum that evolves from a flat two-domain-wall excitation as the transverse field is increased. At the same time, doping generates additional low-energy branches associated with single-domain-wall processes, which become possible due to the presence of holes in the ordered background. Together, these features provide a distinct spectral fingerprint of the hidden-order phase. Across the hidden quantum phase transition, the opening of a spectral gap signals the onset of the paramagnetic phase, where the excitation spectrum is instead dominated by a renormalized Bogoliubov dispersion and broad composite spin-charge continua.

We would like to emphasize that we expect hidden order in other physical settings to show similar spectral signatures, providing information about the respective interplay of local degrees of freedom and ordering with respect to a subspace of the local Hilbert space. 

Since in the doped~\gls{TFIM}, the hidden order phase has the interpretation as gapless~\gls{SPT}, our results have implications for this specific group of phases: We identify the spectral signatures of this exotic phase, in particular signatures of the symmetry protection mechanism in the~\gls{DSF}. 

Taken together, our work establishes dynamical spectroscopy as a practical and powerful probe of hidden order and intrinsically gapless~\gls{SPT} in quantum materials and provides a microscopic understanding of how hidden spin correlations and charge dynamics manifest themselves in experimentally observable spectra.

\section{Acknowledgments}
The authors acknowledge inspiring discussions with Ruben Verresen and Felix Palm. This research was supported by the Deutsche Forschungsgemeinschaft (DFG, German Research Foundation) under Germany’s Excellence Strategy EXC-2111 Grant No. 390814868 and the European Research Council (ERC) under the European Union’s Horizon 2020 research and innovation program (Grant Agreement No. 948141), ERC Starting Grant SimUcQuam and ERC Starting Grant QuaQuaMA (Grant No. 101217531). 
The authors gratefully acknowledge the funding of this project by computing time provided by the Paderborn Center for Parallel Computing ($PC^2$).

\bibliography{references}
\clearpage
\onecolumngrid

\setcounter{page}{1}
\renewcommand{\thepage}{S\arabic{page}}

\setcounter{equation}{0}
\renewcommand{\theequation}{S\arabic{equation}}

\section*{Supplementary Material}

%
%
\section{Numerical DMRG simulations}
\label{sec:appND}
We perform~\gls{DMRG} simulations using the \textsc{SyTen} toolkit to evaluate magnetization $M_S$ ($M_S^*$) in real (squeezed) space.
These are computed from snapshots of the many-body wavefunction via the perfect sampling approach~\cite{ferris_perfect_2012,buser_snapshot-based_2022}.
The sample-averaged magnetization of the state of a system of $L$ sites is computed as
\begin{equation}
    M_S = \frac{1}{N} \sum_i |M_{S,i}|, \quad M_{S,i} = \frac{1}{L} \sum_j (-1)^j S_{j}^z |_i\;.
\end{equation}
Here, $S_{j}^z|_i$ denotes the spin-$z$ value on site $j$ in snapshot $i$ and $N = 10^4$ denotes the total number of sampled snapshots.
In analogy, the sample-averaged squeezed space magnetization is computed with respect to the squeezed space value $S_{j}^z|_{{\rm s},i}$.

\section{Dynamical structure factor simulations}
\label{sec:appSF}
The time evolution of a quantum state is governed by the time-dependent Schr\"odinger equation $i\hbar \partial_t \ket{\psi(t)} = \hat H \ket{\psi (t)}$.
The unitary time evolution operator for an evolution under the Hamiltonian $\hat H$ is given by $U(t) = e^{-it\hat H/\hbar}$ and acts on a state $\ket{\psi(t)}$ via $\hat U(t)\ket{\psi(0)} = \ket{\psi(t)}$.
The~\gls{DSF} is then defined as~\cite{paeckel_time-evolution_2019}
\begin{equation}
\begin{aligned}
		S^{\alpha}(q,\omega) & = \frac{1}{L} \sum_{i}^L \mathrm{e}^{-iqr_i} \int^{\infty}_{-\infty} dt e^{i\omega t} \langle \hat S^\alpha_i(t) \hat S^\alpha_0(0)\rangle_{cc}
		\label{sm:eqtheory:dsf}
\end{aligned}
\end{equation}
where the connected correlator is defined as $\langle \hat S^\alpha_i(t) \hat S^\alpha_0(0)\rangle_{cc} = \bra{\psi_0} \hat U^\dagger(t) \hat S^\alpha_i\hat U(t) \hat S^\alpha_0\ket{\psi_0} - \bra{\psi_0}\hat S^\alpha_i\ket{\psi_0}\bra{\psi_0}\hat S^\alpha_0\ket{\psi_0}$.
The ground states $\ket{\psi_0}$ were obtained by performing~\gls{DMRG} ground-state searches with a maximum bond-dimension of the underlying~\gls{MPS} representations $\chi_\mathrm{max} = 3000$~\cite{white_density-matrix_1993,schollwock_density-matrix_2005,schollwock_density-matrix_2011}.
Instead of computing eq.~\eqref{sm:eqtheory:dsf} directly from a vanilla real-time evolution and subsequent Fourier transformation on a finite time-domain, we employed the recently introduced~\gls{CTKS} expansion~\cite{paeckel_spectral_2026}.
In this approach, the standard technique approximating the infinite-time integration on a finite time domain, which constitutes a major limitation to the frequency resolution in Fig.~\ref{sm:eqtheory:dsf}, is replaced by an exact expression to evaluate the Fourier transform directly in the infinite-time limit.
This is achieved by performing two independent time evolutions of the excited ground states: (i) One real-time evolution until a time $T_\mathrm{real} = N_\mathrm{real}\delta t$ and (ii) one complex-time evolution until a time $T_\mathrm{cplx} = N_\mathrm{cplx} (1-\mathrm i\tan{\alpha})\delta \tau$, where $\delta t,\delta\tau$ denote the chosen time steps and $N_\mathrm{real,cplx}\in\mathbb N$.
These time-evolutions are then combined to construct a~\gls{CTKS}, in which the integral eq.~\eqref{sm:eqtheory:dsf} is evaluated.
Here, we chose $\delta t = 0.1, \delta\tau = 0.5, T_\mathrm{real} = 8 $ and $N_\mathrm{cplx} = 256$ and $\alpha = 0.02\pi$ for $S^z(q,\omega)$ and $\alpha = 0.01\pi$ for $S^x(q,\omega)$.
The time-evolutions were performed using a single-site variant of the~\gls{TDVP}~\cite{haegeman_time-dependent_2011,haegeman_unifying_2016,paeckel_time-evolution_2019} with a local bond expansion scheme~\cite{grundner_cooper-paired_2023}.
Using these tools, we achieve faithful representations of the \gls{DSF} even when using small spectral broadenings $\eta = 0.01$, while keeping the runtimes, which are dominated by the real-time evolution, at an acceptable level.

\begin{figure*}
	\includegraphics[width=\textwidth]{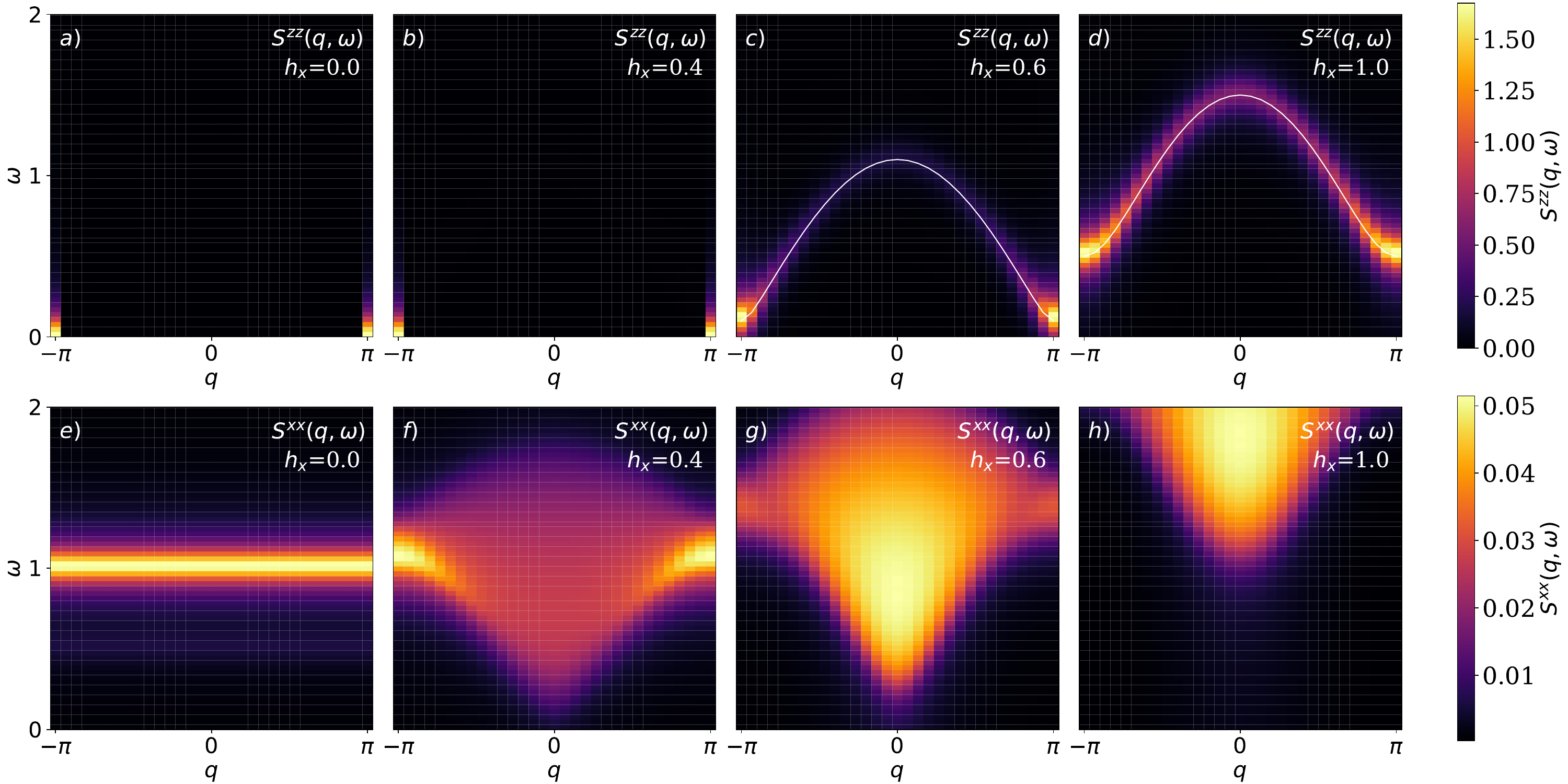}
	\caption{\justifying
		The dynamical spin structure factor $S^{zz}(q,\omega)$ and $S^{xx}(q,\omega)$ of the conventional TFIM for a $h_x$ scan across the \gls{QCP}.
        The numerical data is shown for a system of size $L=30$, and $J =1$.
	\label{fig:dsf_ISING}}
\end{figure*}

For reference, we state the dynamical structure factors of the conventional \gls{TFIM}, $S^{zz}(q,\omega)$ and $S^{xx}(q,\omega)$, see Fig.~\ref{fig:dsf_ISING}.

\section{Squeezed space\label{sec:squeeze}}
In order to investigate the interplay of spin and charge degrees of freedom in the ground state wave function $\vert \psi\rangle =\vert \sigma_1, \dots, \sigma_L\rangle$ with $\sigma_j = \uparrow, \downarrow, h$, we employ the squeezed space formalism:
The Fock state is relabeled as $\vert \psi\rangle = \vert \tilde\sigma_1, \dots, \tilde \sigma_{\tilde L}\rangle \otimes \hat h^\dagger_{x_1}\dots \hat h^\dagger_{x_{N_h}} \vert 0\rangle$, with $\tilde \sigma_j = \uparrow, \downarrow$, $\tilde L = L -N^h$.
The number of holes in the system is given by $N^h = \sum_j n^h_j = \sum_j \hat h^\dagger_j\hat h_j$ and $\hat h^\dagger_j$ corresponds to a hard-core chargon with the same statistics as $\hat c_{j,\sigma}$.
In the following, we denote the hole density as $n^h = N^h/L = 1-n $.
The wave function is hence decoupled into a Fock configuration of spins in squeezed space and a chargon wave function~$\vert \psi\rangle =\vert \psi\rangle_s \otimes\vert \psi\rangle_c$.
The location of a spin in squeezed space is related to a spin in real space via $\tilde j = j - \sum_{l<j} n^h_l$.

Now consider the $t_1-$~\gls{TFIM} in $1D$,
\begin{equation}
\begin{aligned}
		\hat{H} = -t_1 \sum_{\langle i,j\rangle, \sigma} \hat{\mathcal{P}}_{GW}(\hat c^\dagger_{i\sigma}\hat c^{\phantom\dagger}_{j\sigma} + h.c.) \hat{\mathcal{P}}_{GW}
        + J \sum_{\langle i,j\rangle} \hat S^z_i  \hat S^z_j + h_x \sum_j \hat S^x_j \;.
		\label{sm:eqdoped_hopping_TFIM}
\end{aligned}
\end{equation}
We make the Ansatz that eq.~\eqref{sm:eqdoped_hopping_TFIM} takes the following form in squeezed space $\hat H = \hat H_{s} \otimes \mathds{1}_c + \mathds{1}_s \otimes \hat H_{c}$. The individual models are defined as
\begin{equation}
    \hat H_c = -t \sum_{\langle ij\rangle} (\hat h^\dagger_i \hat h_j + \mathrm{h.c})\;,\quad
    \hat H_s = \tilde J  \sum_{\langle \tilde{i}\tilde{j}\rangle} \hat \sigma^z_{\tilde{i}}  \hat \sigma^z_{\tilde{j}} + \tilde h  \sum_{\tilde{j}} \hat\sigma^x_{\tilde{j}}\;.
    \label{sm:eqHcHs}
\end{equation}
The amplitudes are defined as $\tilde J = \frac{J_{\rm eff}}{4}$, $\tilde h = \frac{h_x}{2}$, $t=t_1$ and $J_{\rm eff}$ is introduced below. Spin and Pauli operators are related via $\hat S^\alpha_i = \hat \sigma_i^\alpha/2$.

\subsection{Effective Ising interaction}
\label{sec:appB}
In real space, an Ising interaction only takes place if two neighboring sites $i, i+1$ are occupied.
The effective hopping in squeezed space can hence be derived via the normalized likelihood of finding two neighboring real space sites to be occupied~\cite{hilker_revealing_2017}
\begin{equation}
    J_{\rm eff} = \frac{J}{L-1} \sum_{i=1}^{L-1}\frac{\langle (1-\hat n^h_i)(1-\hat n^h_{i+1})\rangle}{\langle (1-\hat n^h_i)\rangle} \leq J.
    \label{sm:eqJeff}
\end{equation}
The presence of holes leads to an effective Ising interaction $J_{\rm eff} < J$.
For $\vert i-j\vert \rightarrow\infty$ we have $\langle (1-n^h_i)(1-n^h_{j})\rangle \rightarrow \langle (1-n^h_i)\rangle \langle(1-n^h_{j})\rangle$ and $J_{\rm eff}(n) = J  \langle(1-n^h_{j})\rangle = J n$, i.e. the Ising interaction is rescaled by the particle density.
Below, we derive the time dependent spin and charge expectation values required to evaluate eq.~\eqref{sm:eqtheory:dsf}.

\section{Derivation of DSF in squeezed space}
The goal of this section is to analytically compute the~\gls{DSF} eq.~\eqref{sm:eqtheory:dsf} using the squeezed space methods introduced in~\ref{sec:squeeze}. 
For clarity, a summary of definitions can be found in~\ref{sec:defs}.

\subsection{Transforming DSF to squeezed space}
First, we transform the spin operators of the $t_1$-TFIM to squeezed space, $\langle \cdot \rangle \rightarrow \langle \cdot \rangle_s \langle \cdot \rangle_c$.
Explicitly accounting for the hole degree of freedom $ \hat S^\alpha_i \rightarrow \hat S^\alpha_i \hat n_i$ allows us to reduce the local spin Hilbert space dimension from three ($S^\alpha_i = \pm \frac{1}{2},h$) to two ($S^\alpha_i = \pm \frac{1}{2}$).
As a result, the expectation value decouples into a product
\begin{equation}
\begin{aligned}
    \langle \hat S^\alpha_i (t) \hat S^\alpha_0(0)\rangle = \langle \hat S^\alpha_{i}(t) \hat n_{i}(t) \hat S^\alpha_{0}\hat n_{0}\rangle \approx \langle \hat S^\alpha_{\tilde i}(t) \hat S^\alpha_{0}\rangle_s  \langle \hat n_{i}(t) \hat n_{0}\rangle_c\;.
\end{aligned}
\end{equation}
Note that the holes of the original model $\hat n$ corresponds to the charges in squeezed space.
The unconnected part of the correlator eq.~\eqref{sm:eqtheory:dsf} transforms as
\begin{equation}
\begin{aligned}
\langle\hat S^\alpha_i(0)\rangle \langle \hat S^\alpha_0(0)\rangle = \langle\hat S^\alpha_i(0) \hat n_{i}(0)\rangle \langle \hat S^\alpha_0(0)\hat n_{j}(0)\rangle \approx \langle\hat S^\alpha_{\tilde i}(0)\rangle_s \langle \hat S^\alpha_{0}(0)\rangle_s \langle \hat n_{i}(0)\rangle_c \langle\hat n_{0}(0)\rangle_c\;.
\end{aligned}
\end{equation}
Altogether, we have
\begin{equation}
\begin{aligned}
\boxed{
\langle \hat S^\alpha_i(t) \hat S^\alpha_0(0)\rangle_{cc} \rightarrow \langle \hat S^\alpha_{\tilde i}(t) \hat S^\alpha_{0}(0)\rangle_s  \langle \hat n_{i}(t) \hat n_{0}(0)\rangle_c -  \langle\hat S^\alpha_{\tilde i}(0)\rangle_s \langle \hat S^\alpha_{0}(0)\rangle_s \langle \hat n_{i}(0)\rangle_c \langle\hat n_{0}(0)\rangle_c 
}\;.
\label{sm:eqsq_sp_corr}
\end{aligned}
\end{equation}

To evaluate eq.~\eqref{sm:eqtheory:dsf} analytically, we introduce the continuum variable $x = ai$ so that $\tilde{i} \rightarrow xn/a$, with density $n=N/L < 1$, and transform the sum $\sum_{i} \rightarrow \frac{1}{a} \int dx$.
Crucially, in the following we will make the assumption that real and squeezed space sites are related via $\tilde{i} = in$.
The quantity of interest then becomes
\begin{equation}
\begin{aligned}
    S^{\alpha}_{P}(q,\omega) &= \frac{1}{aL} \int^{\infty}_{-\infty} dx \int^{\infty}_{-\infty} dt e^{-i(qx - \omega t)} \langle \hat S^\alpha(x,t) \hat S^\alpha(0,0)\rangle_{cc}\\
    &\approx\frac{a}{4L} \int^{\infty}_{-\infty} dx \int^{\infty}_{-\infty} dt e^{-i(qx - \omega t)} [\langle \hat \sigma^\alpha(xn,t) \hat \sigma^\alpha(0,0)\rangle_s  \langle \hat \rho(x,t) \hat \rho(0,0)\rangle_c - \\
    &\langle\hat \sigma^\alpha(xn,0)\rangle_s \langle \hat \sigma^\alpha(0,0)\rangle_s \langle \hat \rho(x,0)\rangle_c \langle\hat \rho(0,0)\rangle_c]\;.
    \label{sm:eqsf_squeezed}
\end{aligned}
\end{equation}
In the last step -- in order to highlight the distinction between conventional and squeezed space as well as lattice and continuum -- we transform from spins to Paulis $\hat S^\alpha_j \rightarrow \hat\sigma^\alpha(x)/2$ and from number operator to density $\hat n_j/a \rightarrow \hat\rho(x)$, where $N = \sum_j \hat n_j = \int dx \hat\rho (x)$.
Note the density dependent scaling of the space coordinate, which is the origin of several spectral features that we are observing.
We make use of the convolution theorem eq.~\eqref{sm:eqconvolution} to obtain
\begin{equation}
\begin{aligned}
    S^{\alpha}_{P}(q,\omega)\approx  \frac{a}{4L}\int^{\infty}_{-\infty} dt e^{i\omega t} & \overbrace{\mathcal{F}\{\langle \hat \sigma^\alpha(xn,t) \hat \sigma^\alpha(0,0)\rangle_s \}}^{\frac{1}{n}S^{\alpha, \rm TFIM}_P(q/n,t)} * \overbrace{\mathcal{F} \{ \langle \hat \rho(x,t) \hat \rho(0,0)\rangle_c \}}^{S_\rho(q,t)} -  \\
    & \mathcal{F} \underbrace{\{\langle\hat \sigma^\alpha(xn,0)\rangle_s \langle\hat \sigma^\alpha(0,0)\rangle_s\}}_{\frac{1}{n}S^{\alpha, TFIM}_{P,\rm stat}(q/n)} * \underbrace{\mathcal{F} \{\langle \hat \rho(x,0)\rangle_c \langle \hat \rho(0,0)\rangle_c}_{S_{\rho,\rm stat}(q)} \}\\
    =\frac{a}{4nL}\int^{\infty}_{-\infty} dt e^{i\omega t} \int^{\infty}_{-\infty} dk & [ S^{\alpha, \rm TFIM}_{P}(k/n,t) S_\rho(q-k,t) - S^{\alpha, \rm TFIM}_{P,\rm stat}(k/n) S_{\rho,\rm stat}(q-k) ]\;.
    \label{sm:eqsqsp_convolution}
\end{aligned}
\end{equation}
We find that the~\gls{DSF} can be expressed as a momentum-space convolution of the structure factors of the individual spin and charge models. For simplicity we neglected the prefactors below and in the main text.

\subsection{Dynamical structure in the Hidden Order Phase}
Before continuing with our continuum results, we note that in the hidden order phase, we can evaluate the~\gls{DSF} on the lattice via a mean field approach:
In the ground state $\vert 0\rangle_s$ of the hidden order phase, the spins exhibit antiferromagnetic order in squeezed space, i.e. it holds that $\hat S_{\tilde j}^z |0\rangle_s = e^{i\pi \tilde{j}} |0\rangle_s$. Hence, eq.~\eqref{sm:eqsq_sp_corr} becomes 
$e^{i\pi \tilde{i}} [ \langle \hat n_{i}(t) \hat n_{0}(0)\rangle_c - \langle \hat n_{i}(0)\rangle_c \langle\hat n_{0}(0)\rangle_c ]$.
It follows that in this limit the longitudinal dynamical structure factor becomes equivalent to a momentum-shifted charge structure factor.

\begin{equation}
\begin{aligned}
		S^{z}_{HO}(q,\omega) & = \frac{1}{L} \sum_{i}^L \mathrm{e}^{-i(q-\pi n ) r_i} \int^{\infty}_{-\infty} dt e^{i\omega t} \langle \hat n_i(t) \hat n_0(0)\rangle_{cc} \equiv S^{n}(q-\pi n,\omega),
\end{aligned}
\end{equation}
The continuum limit of this expression, $S_\rho (q-\pi n,\omega)$, is derived below, see eq.~\eqref{sm:eqFT_charge_corr}.

\section{Exact solution and DSF of individual models}
\subsection{Tight binding chain}
The tight binding chain eq.~\eqref{sm:eqHcHs} can be diagonalized via a Fourier transform yielding $\hat H_c =\sum_k \varepsilon^c_k \hat h^\dagger_k \hat h_k$ with the lattice dispersion $\varepsilon^c_k = -2t\cos(ak)$.
Writing $\hat n_j^h=\hat h_j^\dagger\hat h_j$ and
$\hat n_j=1-\hat n_j^h$, the connected charge and hole density
correlators coincide. Wick's theorem gives
\begin{equation}
    \langle \hat n_j(t)\hat n_0(0)\rangle
    - \langle\hat n_j\rangle^2
    =
    \langle \hat h_j^\dagger(t)\hat h_0(0)\rangle
    \langle \hat h_j(t)\hat h_0^\dagger(0)\rangle .
\end{equation}
At zero temperature, the two contractions are
\begin{equation}
\begin{aligned}
    \langle \hat h_j^\dagger(t)\hat h_0(0)\rangle
    &= a\int_{\rm BZ}\frac{dk}{2\pi}\,
       n_k e^{-ikja+i\varepsilon_k^c t},\\
    \langle \hat h_j(t)\hat h_0^\dagger(0)\rangle
    &= a\int_{\rm BZ}\frac{dk}{2\pi}\,
       (1-n_k)e^{ikja-i\varepsilon_k^c t},
    \qquad
    n_k=\Theta(k_F^c-|k|).
\end{aligned}
\end{equation}
For the continuum description, we linearize the dispersion
near the two Fermi points,
\begin{equation}
    \varepsilon_{\pm k_F^c+p}^c-\varepsilon_F^c
    \simeq \pm v_F^c p,
    \qquad
    v_F^c=2ta\sin(ak_F^c),
\end{equation}
where $k_F^c=\pi n^h/a$ for hole filling $n^h$.
Keeping the smooth, long-wavelength density component yields
\begin{equation}
\boxed{
    \langle \hat\rho(x,t)\hat\rho(0,0)\rangle
    \simeq \rho_0^2
    -\frac{1}{4\pi^2}
    \left[
        \frac{1}{(x-v_F^c t+i0^+)^2}
        +\frac{1}{(x+v_F^c t-i0^+)^2}
    \right].
}
\end{equation}
Here $\rho_0=\langle\hat n_j\rangle/a=N/(aL)$ for a chain
of $L$ sites. Oscillatory contributions at momentum $2k_F^c$
and corrections from dispersion curvature are neglected.
Away from $x=\pm v_F^c t$, this reduces to
\begin{equation}
    \langle \hat\rho(x,t)\hat\rho(0,0)\rangle
    \simeq
    \rho_0^2-\frac{1}{2\pi^2}
    \frac{x^2+(v_F^c t)^2}
         {[x^2-(v_F^c t)^2]^2}.
\end{equation}
This is the continuum result for the density-density correlator of spinless free fermions in one dimension, as captured by the charge Hamiltonian eq.~\eqref{sm:eqHcHs}.
\subsubsection*{Charge Structure Factor}
To evaluate eq.~\eqref{sm:eqsqsp_convolution}, we define
\begin{equation}
    S_\rho(q,t)
    =\int dx\,e^{-iqx}
      \langle\hat\rho(x,t)\hat\rho(0,0)\rangle.
\end{equation}
The disconnected uniform background contributes
\begin{equation}
    S_{\rho,\mathrm{stat}}(q)
    =\int dx\,e^{-iqx}\rho_0^2
    =2\pi\rho_0^2\delta(q).
\end{equation}
Fourier transforming the smooth connected correlator gives,
in the long-wavelength limit,
\begin{equation}
\boxed{
    S_\rho(q,t)\simeq
    \underbrace{2\pi\rho_0^2\delta(q)}_{S_\rho^0(q)}
    +
    \underbrace{\frac{|q|}{2\pi}
    e^{-i|q|v_F^c t}}_{S_\rho^1(q,t)} .
}
\label{sm:eqFT_charge_corr}
\end{equation}
The first term is the disconnected static background,
while the second describes long-wavelength density
excitations propagating with velocity $v_F^c$.
Contributions near $q=\pm2k_F^c$ are omitted.

\subsection{TFIM\label{sec:TFIM}}
Following standard textbook calculations~\cite{sachdev_quantum_2011}, the 1D Transverse Field Ising model can be solved by mapping the spins to spinless fermions via a Jordan-Wigner transformation:
\begin{equation}
\hat\sigma_j^x = 1-2\hat c_j^\dagger\hat c_j\;,\qquad
\hat\sigma_j^z\hat\sigma_{j+1}^z=-(\hat c_j^\dagger-\hat c_j)(\hat c_{j+1}+\hat c_{j+1}^\dagger)\;.
\label{sm:eqJW}
\end{equation}
Substituting eq.~\eqref{sm:eqJW} into eq.~\eqref{sm:eqHcHs} and dropping the constant $\tilde hL$:
\begin{equation}
\hat H_s = -\tilde J\sum_j(\hat c_j^\dagger-\hat c_j)(\hat c_{j+1}+\hat c_{j+1}^\dagger) - 2\tilde h\sum_j\hat c_j^\dagger\hat c_j\;.
\end{equation}
Expanding exactly as before, using the fermion commutation relations $\{\hat c_j,\hat c_{j+1}^\dagger\}=\{\hat c_j,\hat c_{j+1}\}=0$
\begin{equation}
\boxed{\hat H_s = -\tilde J\sum_j\Big[\hat c_j^\dagger\hat c_{j+1}+\hat c_{j+1}^\dagger\hat c_j+\hat c_j^\dagger\hat c_{j+1}^\dagger+\hat c_{j+1}\hat c_j\Big] - 2\tilde h\sum_j\hat c_j^\dagger\hat c_j}\;.
\label{sm:eqHs_hopping_pairing}
\end{equation}

\textbf{Fourier transform to momentum space.}
With $\hat c_j=\frac{1}{\sqrt L}\sum_k e^{ikja}\hat c_k$, the terms transform as
\begin{equation}
\hat H_s = \sum_k\Big[-2(\tilde J\cos ka+\tilde h)\,\hat c_k^\dagger\hat c_k -i\tilde J\sin(ka)\big(\hat c_k^\dagger\hat c_{-k}^\dagger+\hat c_k\hat c_{-k}\big)\Big]\;.
\end{equation}
Defining $A_k \equiv -2(\tilde J\cos(ka)+\tilde h)\;,B_k \equiv -2\tilde J\sin(ka)$
this takes the form
\begin{equation}
\boxed{\hat H_s = \sum_k\Big[A_k\,\hat c_k^\dagger\hat c_k + \frac{i}{2}B_k\big(\hat c_k^\dagger\hat c_{-k}^\dagger+\hat c_k\hat c_{-k}\big)\Big]}\;.
\label{sm:eqHk}
\end{equation}

\textbf{Diagonalization, angle relations, and dispersion.}
We make the Ansatz
\begin{equation}
\hat c_k = \cos\theta_k\,\hat\gamma_k - i\sin\theta_k\,\hat\gamma_{-k}^\dagger\;,\qquad \theta_{-k}=-\theta_k\;,
\label{sm:eqbogo}
\end{equation}
which is unitary for any real $\theta_k$. Substituting into eq.~\eqref{sm:eqHk} and requiring the terms $\hat\gamma_k^\dagger\hat\gamma_{-k}^\dagger$, $\hat\gamma_{-k}\hat\gamma_k$ to cancel fixes
\begin{equation}
\tan(2\theta_k)=\frac{B_k}{A_k}\;,\qquad \cos(2\theta_k)=\frac{A_k}{\varepsilon_k^s}\;,\qquad \sin(2\theta_k)=\frac{B_k}{\varepsilon_k^s}\;,
\end{equation}
with $\varepsilon_k^s$ fixed by $\cos^2(2\theta_k)+\sin^2(2\theta_k)=1$:
\begin{equation}
\boxed{\hat H_s = \sum_k\varepsilon_k^s\Big(\hat\gamma_k^\dagger\hat\gamma_k-\frac12\Big)\;,\qquad
\varepsilon_k^s = \sqrt{A_k^2+B_k^2} = 2\sqrt{\tilde J^2+\tilde h^2+2\tilde J\tilde h\cos(ka)}}\;.
\label{sm:eqepsk}
\end{equation}
$\varepsilon_k^s$ is minimized where $\cos(ka)$ is most negative, i.e.\ at $k=\pi/a$, giving $\varepsilon^s_{\pi/a}=2|\tilde J-\tilde h|\to0$ at criticality.

\textbf{Fermionic operators in terms of Majoranas.}
With $\hat\gamma_{A,j}=\hat c_j+\hat c_j^\dagger$, $\hat\gamma_{B,j}=-i(\hat c_j-\hat c_j^\dagger)$, i.e.\ $\hat c_j=\tfrac12(\hat\gamma_{A,j}+i\hat\gamma_{B,j})$, Fourier transforming term by term gives
\begin{equation}
\hat c_k = \frac12\big(\hat\gamma_{A,k}+i\hat\gamma_{B,k}\big)\;,\qquad \hat\gamma_{A,k}^\dagger=\hat\gamma_{A,-k}\;,\quad\hat\gamma_{B,k}^\dagger=\hat\gamma_{B,-k}\;.
\end{equation}

\textbf{Occupation number.}
From eq.~\eqref{sm:eqbogo}, $\hat c_k^\dagger=\cos\theta_k\hat\gamma_k^\dagger+i\sin\theta_k\hat\gamma_{-k}$; evaluating in the quasiparticle vacuum $\hat\gamma_k\ket0=0\ \forall k$ (cross terms annihilate $\ket0$ from one side each, $\bra0\hat\gamma_k^\dagger\hat\gamma_k\ket0=0$, $\bra0\hat\gamma_{-k}\hat\gamma_{-k}^\dagger\ket0=1$):
\begin{equation}
\langle\hat c_k^\dagger\hat c_k\rangle = \sin^2\theta_k = \frac12\Big(1-\frac{A_k}{\varepsilon_k^s}\Big)\;.
\end{equation}
Substituting $A_k=-2(\tilde J\cos ka+\tilde h)$:
\begin{equation}
\boxed{\langle\hat c_k^\dagger\hat c_k\rangle = \frac12+\frac{\tilde J\cos(ka)+\tilde h}{\varepsilon_k^s}}\;.
\label{sm:eqoccupation}
\end{equation}
This can then be used to compute the transverse spin expectation value via the Jordan-Wigner relation $\langle \hat{\sigma}^x_j \rangle = 1 - 2\langle \hat{c}^\dagger_j \hat{c}_j \rangle$.

\textbf{Dispersion in the continuum limit.}
To describe the low-energy physics properly away from strict criticality, we expand the exact dispersion relation near the gap minimum at $k = \pi/a$. Letting $k = \pi/a + q$ for small momentum $q$, we expand the cosine as $\cos(ka) = -\cos(qa) \approx -1 + \frac{1}{2}(qa)^2$. Substituting this into the exact dispersion yields
\begin{equation}
\Omega_q^s \approx 2\sqrt{(\tilde{h} - \tilde{J})^2 + \tilde{J}\tilde{h} a^2 q^2}\;.
\end{equation}
Comparing this to the standard relativistic dispersion of a massive Majorana fermion, $E = \sqrt{m^2 + v^2 q^2}$, we can definitively identify the exact mass gap and excitation velocity:
\begin{align}
\boxed{m = 2|\tilde{h} - \tilde{J}|\;, \quad v^s_F = 2a\sqrt{\tilde{J}\tilde{h}}}\;.
\end{align}

\textbf{Chiral fields in momentum space.}
For later convenience we introduce the chiral fields in momentum space as
\begin{equation}
\gamma_A(k)=\frac{1}{\sqrt2}\big(\psi_R(k)+\psi_L(k)\big)\;,\qquad
\gamma_B(k)=\frac{1}{\sqrt2}\big(\psi_R(k)-\psi_L(k)\big)\;.
\end{equation}
Inverting the expressions leads to
\begin{equation}
\boxed{\psi_R(k)=\frac{1}{\sqrt2}\big(\gamma_A(k)+\gamma_B(k)\big)\;,\qquad
\psi_L(k)=\frac{1}{\sqrt2}\big(\gamma_A(k)-\gamma_B(k)\big)}\;.
\end{equation}

\section{Dynamical spin structure factors of individual models}
\subsection{Spin-z structure factor -- Ordered Phase}

The ordered and disordered phases of the \gls{TFIM} require distinct analytical approaches. To account for this in our notation, we introduce an additional index $P$ to $S^{z,\rm{TFIM}}_{P}(q,\omega)$, denoting the ordered (O) and \gls{PM} phase of the \gls{TFIM}.
In order to evaluate eq.~\eqref{sm:eqsqsp_convolution}, we compute $S^{z,\rm{TFIM}}_{O}(k/n,t)$ and $S^{z,\rm{TFIM}}_{O,\rm stat}(k/n)$.
The latter evaluates to
\begin{equation}
    \boxed{S^{z,\rm{TFIM}}_{O,\rm stat}(q/n) = \mathcal{F} \{\langle\hat \sigma^z(xn,0)\rangle_s \langle\hat \sigma^z(0,0)\rangle_s\} = \tilde{M}_{OY}^2 \mathcal{F} \{\cos{(Q_nx)}\} = \pi \tilde{M}_{OY}^2 \left[ \delta\left(q - Q_n \right) + \delta\left(q + Q_n \right)\right]}\;.
\end{equation}
In the above, we have used $\langle \hat \sigma(xn,0)\rangle \langle \hat \sigma(0,0)\rangle = \tilde{M}_{OY}^2 \cos{(Q_nx)}$ with $Q_n = \frac{n\pi}{a}$ and the Onsager-Yang formula $\tilde{M}_{OY} =\langle\hat \sigma^z(0,0)\rangle_s = [1-(\frac{\tilde{h}}{\tilde{J}})^2]^{1/8}$.
Furthermore, in the thermodynamic limit we have
\begin{equation}
\begin{aligned}
    \lim_{x\to\infty}\langle\hat \sigma^z(xn,t)\hat \sigma^z(0,0)\rangle = \langle\hat \sigma^z(xn,t)\rangle \langle\hat \sigma^z(0,0)\rangle\;.
\end{aligned}
\end{equation}
Performing the Fourier transform then leads to
\begin{equation}
    \boxed{S^{z,\rm{TFIM}}_O (q/n,t)  = \mathcal{F}\{\langle \hat \sigma^z(xn,t) \hat \sigma^z(0,0)\rangle_s\} = \pi \tilde{M}_{OY}^2 \left[ \delta\left(q - Q_n \right) + \delta\left(q + Q_n \right) \right]  = S^{z,\rm{TFIM}}_{O,\rm stat} (q)} \;.
    \label{sm:eqFT_spin_ordered_corr}
\end{equation}
$S^{z,\rm{TFIM}}_{O} (q,t)$ can alternatively be computed via a form factor expansion (see below).
However, due to the observation that higher orders of this expansion do not contribute dominant spectral features in the regime of interest, they are being neglected and we proceed with the dominant contribution obtained above.
Furthermore note that we are performing a convolution of spin and charge degrees of freedom.
Hence, the above equality does not lead to a vanishing of the full structure factor.

\subsection{Spin-z structure factor -- Paramagnetic Phase}

As for the disordered phase, we perform a form factor expansion.
The two-point correlation function is dominated by the 1-particle sector. 
The Lehmann representation is given by
\begin{equation}
C^{PM}_\sigma(x, t) = \langle\hat \sigma^z(x,t)\hat \sigma^z(0,0)\rangle = \int \frac{dP}{2\pi \cdot 2E} |F_1|^2 e^{i(Px - Et)}\;,
\end{equation}
where $E = \sqrt{m^2 + (v^s_F)^2 P^2}$ and $|F_1|^2 = \bar{s}^2$ is the constant form factor in the scaling limit.
To solve the integral, we introduce the rapidity $\theta$, such that
\begin{equation}
E = m \cosh \theta, \quad P = \frac{m}{v^s_F} \sinh \theta, \quad dP = \frac{m}{v^s_F} \cosh \theta d\theta\;.
\end{equation}
Substituting these into the spectral integral, the measure simplifies significantly ($dP/E = (v^s_F)^{-1} d\theta$) yielding
\begin{equation}
C^{PM}_\sigma(x, t) = \frac{\bar{s}^2}{4\pi v^s_F} \int_{-\infty}^{\infty} d\theta \exp\left[ i \frac{m}{v^s_F} \left( x \sinh \theta - v^s_F t \cosh \theta \right) \right]\;.
\end{equation}
We define $s = \sqrt{x^2 - (v^s_F t)^2}$. By performing a Wick rotation $t \to -i\tau$, the exponent takes the form of a standard hyperbolic integral
\begin{align}
C^{PM}_\sigma(x, \tau) & = \frac{\bar{s}^2}{4\pi v^s_F} \int_{-\infty}^{\infty} d\theta \exp\left[ -\frac{m}{v^s_F} \sqrt{x^2 + (v^s_F)^2 \tau^2} \cosh(\theta - \phi) \right]\\
&= \frac{\bar{s}^2}{2\pi v^s_F} \mathcal{K}_0\left( \frac{m}{v^s_F} \sqrt{x^2 - (v^s_F t)^2} \right)\;.
\end{align}
In the last step we used the integral identity for the modified Bessel function $\mathcal{K}_0(z) = \frac{1}{2} \int_{-\infty}^{\infty} e^{-z \cosh \theta} d\theta$ we obtain the final real-space analytical expression
However, in the \gls{AFM} TFIM, the order parameter is staggered.
Hence, we account for the modulation of the correlation function by the N\'eel vector $Q = \pi/a$ by inserting an oscillating prefactor.
The Fourier transform is then given by
\begin{equation}
S^{PM}_\sigma(q, t) = \frac{\bar{s}^2}{2\pi v^s_F} \int_{-\infty}^{\infty} dx \, e^{-i(q-Q)x} \mathcal{K}_0\left( \frac{m}{v^s_F} \sqrt{x^2 - v_F^2 t^2} \right)\;.
\end{equation}
Using the Bessel integral identity, we perform the transformation:
\begin{equation}
\int_{-\infty}^{\infty} dx \, e^{-ipx} \mathcal{K}_0(\frac{m}{v^s_F}\sqrt{x^2 - (v^s_F)^2 t^2}) = \frac{\pi v^s_F \cos(t \sqrt{m^2 + (v^s_F)^2 p^2})}{\sqrt{m^2 + (v^s_F)^2 p^2}}\;.
\end{equation}
This leads directly to the final form for the disordered AFM phase
\begin{equation}
\boxed{S^{z,\rm{TFIM}}_{PM} = \frac{\bar{s}^2}{2 \Omega^{s,\pm}_q} \cos(\Omega^{s,\pm}_q t)\;, \quad \Omega^{s,\pm}_q = \sqrt{m^2 + (v^s_F)^2 (q \pm \pi/a)^2}}\;.
\label{sm:eqFT_spin_disordered_corr}
\end{equation}
with $\bar s = \abs{F_1}^2 \propto m^{1/4}$, the absolute square of the one-particle form factor~\cite{wu_spin-spin_1976}.

\subsection{Spin-x structure factor}

The disconnected part of the transverse structure factor isolates the uniform background magnetization:
\begin{equation}
\boxed{S^{x,\rm{TFIM}}_{\rm stat}(q) = \mathcal{F} \{\langle\hat \sigma^x(x)\rangle \langle\hat \sigma^x(0)\rangle\}= 2\pi \langle \sigma^x \rangle^2  \delta(q) = S^{x, \rm TFIM}_{\rm stat}(q)}\;.
\end{equation}
This indicates an elastic peak exactly at zero momentum transfer ($q=0$).
The transverse spin operator maps to the energy density operator, $\hat\sigma^x(x)\to\epsilon(x)=i\psi_R(x)\psi_L(x)$, which in momentum space -- using $\psi_{R,L}(k)=\tfrac1{\sqrt2}(\gamma_A(k)\pm\gamma_B(k))$ -- reads
\begin{equation}
\epsilon(q) = i\int\frac{dk}{2\pi}\,\psi_R(k)\,\psi_L(q-k)\;.
\end{equation}

\textbf{Bogoliubov substitution.} Diagonalizing $\hat{\mathcal H}_{\rm Ising}$ introduces continuum quasiparticle operators $\hat\gamma_k$ via
\begin{equation}
\psi_R(k) = \cos\theta_k\,\hat\gamma_k + i\sin\theta_k\,\hat\gamma_{-k}^\dagger\;,\qquad
\psi_L(k) = i\sin\theta_k\,\hat\gamma_k + \cos\theta_k\,\hat\gamma_{-k}^\dagger\;.
\end{equation}
For the angle, we keep the same pairing established for the lattice transformation in~\ref{sec:TFIM} ($\cos2\theta_k=A_k/\varepsilon_k$ with $A_k$ the mass-like/diagonal term, $\sin2\theta_k=B_k/\varepsilon_k^s$ with $B_k$ the momentum-like/off-diagonal term) --- consistent with linearizing $A_k\to \pm m$, $B_k\to v_F^sk$ near the gap point:
\begin{equation}
\cos(2\theta_k) = \frac{m}{\Omega_k^s}\;,\qquad \sin(2\theta_k) = \frac{v_F^sk}{\Omega_k^s}\;,\qquad \Omega_k^s=\sqrt{m^2+(v_F^s)^2k^2}\;.
\end{equation}
Since $\hat\gamma_k\ket0=0$, only the creation-operator pieces of $\psi_R(k),\psi_L(q-k)$ contribute to $\epsilon(q)\ket0$; the amplitude to create quasiparticles at $-k$ and $k-q$ is $A(k,q)=\sin\theta_k\cos\theta_{q-k}$. Because the created Majorana quasiparticles anticommute, the physical amplitude is the antisymmetrized combination
\begin{equation}
\mathcal M(k,q) = A(k,q)-A(q-k,q) = \sin\theta_k\cos\theta_{q-k}-\cos\theta_k\sin\theta_{q-k} = \sin(\theta_k-\theta_{q-k})\;.
\end{equation}
Squaring, using the angle-subtraction identity:
\begin{align}
|\mathcal M(k,q)|^2 &= \sin^2(\theta_k-\theta_{q-k}) = \frac12\Big[1-\cos(2\theta_k-2\theta_{q-k})\Big]\nonumber\\
&=\frac12\Big[1-\big(\cos2\theta_k\cos2\theta_{q-k}+\sin2\theta_k\sin2\theta_{q-k}\big)\Big]
=\frac12\left(1-\frac{m^2+(v_F^s)^2k(q-k)}{\Omega_k^s\,\Omega_{q-k}^s}\right)\;.
\end{align}

The connected dynamical structure factor $S_{\rm conn}(q,t)=\langle0|\epsilon(q,t)\epsilon(-q,0)|0\rangle_{\rm conn}$ measures the time evolution of these two created quasiparticles. Combining the phase-space integral, the coherence factor, and the dynamical phase $e^{-i(\Omega_k^s+\Omega_{q-k}^s)t}$:
\begin{equation}
\boxed{S^{x,\rm{TFIM}}_{\rm conn}(q,t) = \int_{-\infty}^\infty\frac{dk}{2\pi}\,\frac12\left(1-\frac{m^2+(v_F^s)^2k(q-k)}{\Omega_k^s\,\Omega_{q-k}^s}\right)e^{-i(\Omega_k^s+\Omega_{q-k}^s)t}}\;.
\end{equation}
The static structure factor is given by 
\begin{equation}
\boxed{S^{x,\rm{TFIM}}_{\rm stat}(q) = \langle\hat\sigma^x\rangle^2\delta(q)}\;.
\end{equation}
with $\langle\sigma^x\rangle$ obtained via eq.~\eqref{sm:eqoccupation}.

\section{Dynamical structure factor $S^{z}_{cc}(q,\omega)$ in squeezed space}
\begin{figure*}
	\includegraphics[width=\textwidth]{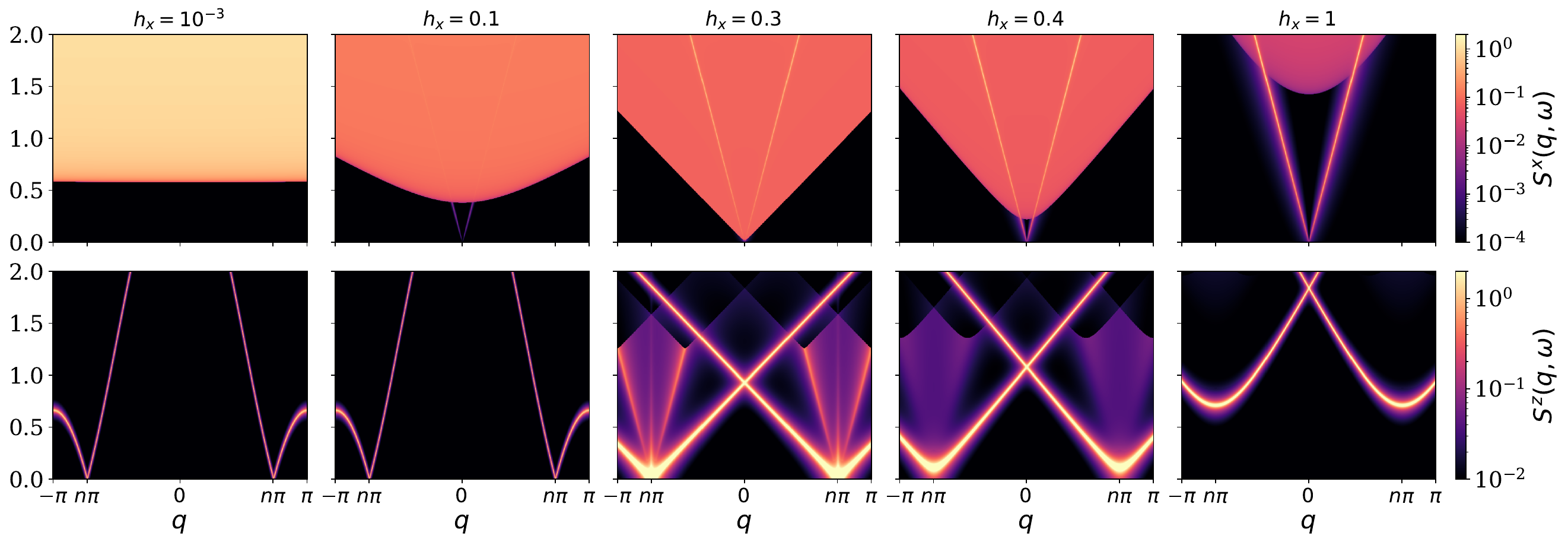}
	\caption{\justifying
		Analytical results for $S^x$ (upper row) and $S^z$ (bottom row) for a $h_x$ scan across the hidden~\gls{QCP}.
        This figure shows the pure continuum treatment without accounting for the $k$-dependence of the dispersions or velocities.
	\label{fig:dsf_continuum}}
\end{figure*}

The below calculations are performed in the continuum which (not only but also) results in linearized dispersions. However, in order to obtain our final results, we replaced the linearized dispersion by the lattice dispersion and took into account the $k$-dependence of the Fermi velocity to account for lattice effects expected in experimental settings. For reference, we plot the final expressions using the full analytical treatment in Fig.~\ref{fig:dsf_continuum}.

\subsection{Ordered phase}
Since the unconnected term is being canceled, we only need to compute
\begin{equation}
\begin{aligned}
S^{z}_{HO}(q,\omega)&= \int^{\infty}_{-\infty} dt e^{i\omega t}\int^{\infty}_{-\infty} dk [ S^{z,\rm{TFIM}}_O (k/n,t) S^1_\rho(q-k,t)]\;,
\end{aligned}
\end{equation}
using $ S^{z,\rm{TFIM}}_O (k,t)$ from eq.~\eqref{sm:eqFT_spin_ordered_corr} and $S^1_\rho(q-k,t)$ from eq.~\eqref{sm:eqFT_charge_corr}.
The convolution in the ordered phase leads to linear dispersive branches
\begin{equation}
\boxed{S^{z}_{HO}(q,\omega) = \pi \tilde{M}_{OY}^2 \left[ |q-Q_n| \delta(\omega - v^c_F |q-Q_n|) + |q+Q_n| \delta(\omega - v^c_F |q+Q_n|) \right]}
\end{equation}
with $Q_n = \frac{n\pi}{a}$. Note that in the main text we restrict our considerations to $\omega > 0$.

\subsection{Paramagnetic phase}
Computing eq.~\eqref{sm:eqsqsp_convolution} in the disordered phase is slightly more involved.
We need to evaluate 
\begin{equation}
\begin{aligned}
S^{z}_{PM}(q,\omega)&= \int^{\infty}_{-\infty} dt e^{i\omega t}\int^{\infty}_{-\infty} dk
    [ S^{z,\rm{TFIM}}_{PM} ((q-k)/n,t) S^0_\rho(k,t) + S^{z,\rm{TFIM}}_{PM} ((q-k)/n,t) S^1_\rho(k,t)]\\
    &= S^{PM,0}(q,\omega) + S^{PM,1}(q,\omega)\;, 
    \label{sm:eqdisorderedSzwk}
\end{aligned}
\end{equation}
using eq.~\eqref{sm:eqFT_spin_disordered_corr} and eq.~\eqref{sm:eqFT_charge_corr}.

\textbf{I. Rescaled spin-wave term $S^{PM,0}(q,\omega)$}.
We start by performing the momentum convolution
\begin{equation}
    \int^{\infty}_{-\infty} dk \; S^{z,\rm{TFIM}}_{PM} ((q-k)/n,t) S^0_\rho(k,t) =
    \int dk \frac{\bar{s}^2 \cos(\Omega^{s,\pm}_{(q-k)/n} t)}{2 \Omega^{s,\pm}_{(q-k)/n}} 2\pi \rho_0^2 \delta(k) = \frac{\pi \rho_0^2 \bar{s}^2}{\Omega^{s,\pm}_{q/n}} \cos(\Omega^{s,\pm}_{q/n} t)\;.
\end{equation}
The time Fourier transform of the cosine yields a pair of Dirac delta functions in frequency space
\begin{equation}
    \boxed{S^{PM,0}(q,\omega) = \frac{\pi^2 \rho_0^2 \bar{s}^2}{\Omega^{s,\pm}_{q/n}} \left[ \delta(\omega - \Omega^{s,\pm}_{q/n}) + \delta(\omega + \Omega^{s,\pm}_{q/n}) \right]\;}.
\end{equation}
This term describes elastic background scattering.

\textbf{II. Composite excitation term $S^{PM,1}(q,\omega)$}.
\begin{equation}
    \int^{\infty}_{-\infty} dk \; S^{z,\rm{TFIM}}_{PM} ((q-k)/n,t) S^1_\rho(k,t)
 = \frac{\bar{s}^2}{8\pi} \int_{-\infty}^{\infty} dk \frac{|k|}{\Omega^{s,\pm}_{(q-k)/n}} \left[ e^{i(\Omega^{s,\pm}_{(q-k)/n} - |k| v_F^c)t} + e^{-i(\Omega^{s,\pm}_{(q-k)/n} + |k| v_F^c)t} \right]\;.
\end{equation}
In this case, we first solve the time integral using $\int e^{i(\omega \pm \Omega - |k|v_c)t} dt = 2 \pi\delta (\omega \pm \Omega - \abs{k}v_c)$ and obtain
\begin{equation}
\boxed{S^{PM,1}(q,\omega)= \frac{\bar{s}^2}{4}\int^{\infty}_{-\infty} dk \frac{\vert k\vert}{\Omega^{s,\pm}_{(q-k)/n}} \left[ \delta(\omega - \Omega^{s,\pm}_{(q-k)/n} - |k|v_F^c) + \delta(\omega + \Omega^{s,\pm}_{(q-k)/n} - |k|v_F^c) \right]\;.}
\label{sm:eqcontinuum_disordered}
\end{equation}

\textbf{Root method for integral evaluation.}
As an example, we use $S^{PM,1}(q,\omega)$ to demonstrate the evaluation of the integral via the roots method which is used in a similar manner for $S^{x}(q,\omega)$.
The threshold energy, $\omega_{th}(q)$, is the absolute minimum energy required to excite this pair of modes for a given momentum transfer $q$. It is defined by the global minimum of the combined excitation energy with respect to the momentum $k$:
\begin{equation}
\omega_{th}(q) = \min_k \left( \Omega^{s,\pm}_{(q-k)/n} + |k|v_F^c \right) = m + |k|v_F^c 
\end{equation}

To numerically evaluate this integral, we utilize the property of the Dirac delta function $\int f(k) \delta(g(k)) dk = \sum_{i} \frac{f(k_i)}{|g'(k_i)|}$ where $k_i$ are the roots such that $g(k_i) = 0$.
We identify 
\begin{equation}
    f(k) = \frac{\vert k\vert }{\Omega^{s,\pm}_{(q-k)/n}}\;, \quad g_\pm(k) = \omega \pm \Omega^{s,\pm}_{(q-k)/n} - |k|v_F^c = 0
\end{equation}
The derivatives with respect to $k$ are $g'_\pm(k) = \frac{d}{dk} \left(\pm\Omega^{s,\pm}_{(q-k)/n} + |k|v_F^c \right)$.
We numerically evaluate
\begin{equation}
S^{PM,1}_{\pm}(q,\omega)=\frac{\bar{s}^2}{4} \sum_{i} \frac{\vert k_i\vert}{\Omega^{s,\pm}_{(q-k_i)/n} \left| \frac{\partial}{\partial k} (\pm \Omega^{s,\pm}_{(q-k)/n} + |k|v_F^c) \right|_{k=k_i}}\;.
\end{equation}

\section{Dynamical structure factor $S^{x}(q,\omega)$ in squeezed space}
In order to evaluate eq.~\eqref{sm:eqsqsp_convolution} for $\alpha = x$ we need to compute
\begin{equation}
\begin{aligned}
S^{x}(q,\omega)
    &= \int^{\infty}_{-\infty} dt e^{i\omega t}\int^{\infty}_{-\infty} dk
    S^{x,\rm{TFIM}}_{\rm{conn}}(q-k,t)[ S^0_\rho(k,t) + S^1_\rho(k,t)] + S^{x,\rm{TFIM}}_{\rm stat} (q-k,t) S^1_\rho(k,t)\\
    &= C_0(q,\omega) + C_1(q,\omega) + C_1^{\rm stat}(q,\omega)
    \label{sm:eqdisorderedSxwk}
\end{aligned}
\end{equation}
Let us evaluate these term by term:

\textbf{I. The Static Background Term $C_0(q,\omega)$.}
Evaluating the convolution for the first term yields:
\begin{equation}
C_0(q,t) = \int \frac{dk}{2\pi} S_{\rm{conn}}^{x,TFIM }((q - k)/n, t) \left[ 2\pi \rho_0^{\,2}\,\delta(k) \right]  = \rho_0^{\,2} S_{\rm{conn}}^{x,\rm{TFIM}}(q/n, t)
\end{equation}
Taking the Fourier transform to the frequency domain ($\omega$) simply reproduces the rescaled Ising connected correlator scaled by a constant amplitude:
\begin{equation}
\boxed{C_0(q,\omega) = \rho_0^{\,2} S^{x,\rm TFIM}_{\rm{conn}}\left(\frac{q}{n}, \omega\right)}
\end{equation}

\textbf{II. The Dynamical Acoustic Fluctuation Term $C_1(q,\omega)$.}
The second term introduces the dynamical fluctuations of the $\rho$ field:
\begin{equation}
C_1(q,t) = \int \frac{dk}{2\pi} S_{\rm{conn}}^{x,\rm TFIM}((q - k)/n, t) \left( \frac{|k|}{2\pi}\,e^{-i |k| v^c_F t} \right) 
\end{equation}
To evaluate this in the frequency domain, we apply the temporal Fourier transform. Exchanging the order of integration gives:
\begin{equation}
C_1(q,\omega) = \int \frac{dk}{(2\pi)^2} \left( |k| \right) \int dt \, e^{i(\omega - |k|v^c_F)t} S_{\rm{conn}}^{x,\rm TFIM}((q - k)/n, t)
\end{equation}
The inner time integral is exactly the Fourier transform of the Ising correlator, evaluated at a shifted frequency $\Omega = \omega - |k|v^c_F$:
\begin{equation}
\boxed{C_1(q,\omega) = \int \frac{dk}{(2\pi)^2} |k| S^{x,\rm TFIM}_{\rm{conn}}\left( \frac{q - k}{n}, \omega - |k|v^c_F \right)}
\end{equation}
\textbf{III. The Static Acoustic Fluctuation Term $C^{\rm stat}_1(q,\omega)$.}
The third term isolates the coupling between the gapless acoustic fluctuations of the charge field and the static part of the spin correlator. We start with the convolution in the time domain:
\begin{equation}
C^{\rm stat}_1(q,t) = \int \frac{dk}{2\pi} S_{\rm stat}^{x, \rm TFIM}((q - k)/n) \left( \frac{|k|}{2\pi}\,e^{-i |k| v^c_F t} \right) 
\end{equation}
Since the static spin structure factor is independent of time and heavily peaked at zero momentum, we can substitute $S_{\rm stat}^{x,\rm TFIM}(q-k, t) = \langle \hat \sigma^x\rangle^2\delta(q-k)$. Evaluating the integral over $k$ yields:
\begin{equation}
C^{\rm stat}_1(q,t) =  \frac{\langle\hat \sigma^x\rangle^2|q|}{(2\pi)^2} e^{-i |q| v^c_F t}
\end{equation}
Taking the Fourier transform to the frequency domain involves computing the time integral $\int dt \, e^{i\omega t}$. This directly evaluates to a delta function
\begin{equation}
\boxed{C^{\rm stat}_1(q,\omega) = \frac{\langle\hat \sigma^x\rangle^2 |q|}{2\pi} \delta(\omega - |q|v^c_F)}\;, 
\end{equation}
where the expectation value is evaluated via eq.~\eqref{sm:eqoccupation}.

\section{Definitions\label{sec:defs}}
The spatial Fourier transform is defined as 
\begin{equation}
    C(q) = \mathcal{F}\{C(x)\} =\int_{-\infty}^{\infty} dx \, e^{-iqx} C(x)\;, \quad C(x) = \mathcal{F}^{-1}\{C(q)\} = \frac{1}{2\pi} \int_{-\infty}^{\infty} dq \, e^{iqx} C(q)\;.
\end{equation}
The temporal Fourier transform is defined as 
\begin{equation}
    C(\omega) = \int_{-\infty}^{\infty} dt \, e^{i\omega t} C(t)\;, \quad C(t) = \frac{1}{2\pi} \int_{-\infty}^{\infty} d\omega \, e^{-i\omega t} C(\omega)\;.
\end{equation}
We furthermore make use of the convolution theorem
\begin{equation}
    \mathcal{F} \{ f(x)\cdot g(x)\} = (F*G)(k), \qquad (F*G)(k) = \int_{-\infty}^{\infty} dq\; F(q)G(k-q)\;,
    \label{sm:eqconvolution}
\end{equation}
with $F(k) = \mathcal{F}\{f(x)\}, G(k) = \mathcal{F}\{g(x)\}$.
Other useful identities include $\int_{-\infty}^{\infty} dx \, e^{-i(q-Q)x} = 2\pi \delta(q-Q)$, $\delta (q/n-Q) = \vert n \vert \delta(q-nQ)$ as well as $\mathcal{F}\{f(nx)\} = \frac{1}{\vert n\vert}\hat f(q/n)$.

\end{document}